\documentclass[iop]{emulateapj}

\newcommand{\cjaa} {ChJAA}

\newcommand{\xmm} {{\it XMM-Newton}}
\newcommand{\chandra} {{\it Chandra}}
\newcommand{\nustar} {{\it NuSTAR}}

\newcommand{\cmsq} {cm$^{-2}$}
\newcommand{\nh} {$N_{\rm{H}}$}
\newcommand{\lx} {$L_{\rm{X}}$}
\newcommand{\fx} {$F_{\rm{X}}$}
\newcommand{\chisq} {$\chi^2$}

\newcommand{\degree}{{$^\circ$}}
\newcommand{\ergs}{\mbox{\thinspace erg\thinspace s$^{-1}$}}
\newcommand{\ergcms}{\mbox{\thinspace erg\thinspace cm$^{-2}$\thinspace s$^{-1}$}}

\newcommand{\mbh} {$M_{\rm BH}$}

\newcommand{\lamedd} {$\lambda_{\rm Edd}$}
\newcommand{\fcov} {$f_{\rm c}$}
\newcommand{\thetator} {$\theta_{\rm tor}$}

\shorttitle{The covering factor of Compton-thick AGN with NuSTAR}
\shortauthors{Brightman et al.}

\begin{document}

\title{Determining the covering factor of Compton-thick active galactic nuclei with NuSTAR}

\author{M. Brightman$^{1,2}$, M. Balokovi\'c$^{1}$, D. Stern$^{3}$, P. Ar\'{e}valo$^{4}$, D. R. Ballantyne$^{5}$, F. E. Bauer$^{6,7,8}$, S. E. Boggs$^{9}$, W. W. Craig$^{9,10}$, F. E. Christensen$^{11}$, A. Comastri$^{12}$, F. Fuerst$^{1}$, P. Gandhi$^{13,14}$, C. J. Hailey$^{15}$, F. A. Harrison$^{1}$, R. C. Hickox$^{16}$, M. Koss$^{17}$, S. LaMassa$^{18}$, S. Puccetti$^{19,20}$, E. Rivers$^{1}$, R. Vasudevan$^{21}$, D. J. Walton$^{3,1}$, and W. W. Zhang$^{22}$}

\affil{$^{1}$Cahill Center for Astrophysics, California Institute of Technology, 1216 East California Boulevard, Pasadena, CA 91125, USA\\
$^{2}$Max-Planck-Institut f\"{u}r extraterrestrische Physik, Giessenbachstrasse 1, D-85748, Garching bei M\"{u}nchen, Germany\\
$^{3}$Jet Propulsion Laboratory, California Institute of Technology, 4800 Oak Grove Drive, Mail Stop 169-221, Pasadena, CA 91109, USA\\
$^{4}$Instituto de F\'{i}sica y Astronoma, Facultad de Ciencias, Universidad de Valpara\'{i}so, Gran Bretana N 1111, Playa Ancha, Valpara\'{i}so, Chile\\
$^{5}$Center for Relativistic Astrophysics, School of Physics, Georgia Institute of Technology, Atlanta, GA 30332, USA\\
$^{6}$Instituto de Astrof\'{i}sica, Facultad de F\'{i}sica, Pontificia Universidad Catolica de Chile, Casilla 306, Santiago 22, Chile\\
$^{7}$Millennium Institute of Astrophysics\\
$^{8}$Space Science Institute, 4750 Walnut Street, Suite 205, Boulder, CO 80301, USA\\
$^{9}$Space Sciences Laboratory, University of California, Berkeley, CA 94720, USA\\
$^{10}$Lawrence Livermore National Laboratory, Livermore, CA 94550, USA\\
$^{11}$DTU Space, National Space Institute, Technical University of Denmark, Elektrovej 327, DK-2800 Lyngby, Denmark\\
$^{12}$INAF Osservatorio Astronomico di Bologna, via Ranzani 1, I-40127 Bologna, Italy\\
$^{13}$Department of Physics, Durham University, South Road, Durham DH1 3LE, UK\\
$^{14}$School of Physics \& Astronomy, University of Southampton, Highfield, Southampton SO17 1BJ\\
$^{15}$Columbia Astrophysics Laboratory, Columbia University, New York, NY 10027, USA\\
$^{16}$Department of Physics and Astronomy, Dartmouth College, 6127 Wilder Laboratory, Hanover, NH 03755, USA\\
$^{17}$SNSF Ambizione Fellow, Institute for Astronomy, Department of Physics, ETH Zurich, Wolfgang-Pauli-Strasse 27, CH-8093 Zurich, Switzerland\\
$^{18}$Yale Center for Astronomy and Astrophysics, Yale University, P.O. Box 208120, New Haven, CT 06520, USA\\
$^{19}$ASDC-ASI, Via del Politecnico, I-00133 Roma, Italy\\
$^{20}$INAF-Osservatorio Astronomico di Roma, via Frascati 33, I-00040 Monte Porzio Catone (RM), Italy\\
$^{21}$Institute of Astronomy, Madingley Road, Cambridge CB3 0HA, UK\\
$^{22}$NASA Goddard Space Flight Center, Greenbelt, MD 20771, USA\\}

\begin{abstract}

The covering factor of Compton-thick obscuring material associated with the torus in active galactic nuclei (AGN) is at present best understood through the fraction of sources exhibiting Compton-thick absorption along the line of sight (\nh$>1.5\times10^{24}$ \cmsq) in the X-ray band, which reveals the average covering factor. Determining this Compton-thick fraction is difficult however, due to the extreme obscuration. With its spectral coverage at hard X-rays ($>$10 keV), \nustar\ is sensitive to the AGN covering factor since Compton scattering of X-rays off optically thick material dominates at these energies. We present a spectral analysis of 10 AGN observed with \nustar\ where the obscuring medium is optically thick to Compton scattering, so called Compton-thick (CT) AGN. We use the torus models of Brightman \& Nandra which predict the X-ray spectrum from reprocessing in a torus and include the torus opening angle as a free parameter and aim to determine the covering factor of the Compton-thick gas in these sources individually. Across the sample we find mild to heavy Compton-thick columns, with \nh\ measured from $10^{24}$--$10^{26}$ \cmsq, and a wide range of covering factors, where individual measurements range from 0.2--0.9. We find that the covering factor, $f_{\rm c}$, is a strongly decreasing function of the intrinsic 2-10 keV luminosity, \lx, where $f_{\rm c}=(-0.41\pm0.13)$log$_{10}$(\lx/\ergs)$+18.31\pm5.33$, across more than two orders of magnitude in \lx\ (10$^{41.5}$--10$^{44}$ \ergs). The covering factors measured here agree well with the obscured fraction as a function of \lx\ as determined by studies of local AGN with \lx$>10^{42.5}$ \ergs. 

\end{abstract}


\section{Introduction}

The average covering factor of the obscurer in active galactic nuclei (AGN), understood to be a torus-like structure, is represented by the ratio of obscured to unobscured AGN, or the obscured fraction. The obscured fraction has been well studied at many wavelengths, such as the optical \citep[e.g.][]{lawrence91, simpson05}, the mid-infrared \citep[e.g.][]{lusso13,gu13,assef13} and the X-rays \citep[e.g.][]{risaliti99} and is largely understood to be dependent on the power of the central source \citep[e.g.][]{ueda03, lafranca05, akylas06, tueller08, hasinger08, beckmann09, akylas09, burlon11, brightman11b, vasudevan13}, although some works have shown that this may be an observational effect \citep[e.g.][]{lawrence10,mayo13}. 

The fraction of AGN where the obscuration is so extreme that it is optically thick to Compton scattering (\nh$>1.5\times10^{24}$ \cmsq), so called Compton-thick (CT) AGN, is less well known \citep[e.g.][]{burlon11,brightman11b}, with only a few tens of bonafide CT AGN known locally \citep[e.g.][]{goulding12,gandhi14}. The dependency of the covering factor of Compton-thick gas on the power of the AGN is relatively unknown, even in the local universe, hindered by low number statistics. At the higher redshifts probed by deep extragalactic surveys by \chandra\ and \xmm, the redshifting of the Compton-hump into the bandpasses of these telescopes aids identification of these sources \citep[e.g.][]{comastri11}. In addition, the large volumes probed yield of order hundreds to thousands of sources overall such that significant numbers of CT AGN can be uncovered \citep[e.g.][]{brightman12b,buchner14,brightman14}. Nevertheless, the low count nature of these sources means that constraints on the \nh\ and \lx\ are still relatively poor.

The ability to determine covering factors for individual sources is needed to solve many of the issues mentioned above and is needed in order to carry out a detailed study of what physically affects the covering factor. Estimating the covering factor in individual sources can be done by determining the ratio of emission reprocessed in the obscuring medium to the intrinsic emission, which is best done in unobscured sources (i.e. those viewed with a clear line of sight to the nucleus) where the intrinsic emission is directly visible \citep[e.g.][]{treister08,gandhi09,lusso13,gu13}. However the intrinsic emission in such sources can dominate over the reprocessed emission, making them difficult to disentangle and furthermore disk reflection is difficult to separate from distant reflection. In obscured sources (i.e. those viewed through thick material) the reprocessed emission dominates, though estimates of the intrinsic emission are challenging due to the obscuration itself. 

X-ray spectral analysis extending to high energies is ideal for determining the covering factor due to the fact that X-rays above $\sim$3 keV penetrate all but the most extreme obscuration, allowing a good estimate of the intrinsic power. Furthermore, in CT AGN where the obscuring medium is optically thick to Compton scattering, the scattering of X-rays within the medium can reveal the covering factor of the gas for such sources, especially evident above 10 keV. This has been done for a handful of local AGN, mostly with the use of {\it Suzaku} data \citep[e.g.][]{awaki09, eguchi11, tazaki11, yaqoob12, kawamuro13}. However, a large statistical analysis on what physically influences the AGN covering factor has yet to be carried out.

The recently launched {\it Nuclear Spectroscopic Telescope Array} \citep[\nustar,][]{harrison13} is sensitive in the 3-79 keV band. Its significantly improved sensitivity above 10 keV with respect to previous telescopes makes it the ideal instrument to measure the strength and shape of the Compton reflection hump and thus determine the covering factor in a large number of AGN. As part of its baseline mission, \nustar\ has observed $\sim$100 AGN from the hard X-ray selected {\it Swift}/BAT survey for $\sim$20 ks each, as well as longer observations of several well known obscured AGN such as NGC~1068 and NGC~4945. This sample, which has also been studied at many other wavelengths, is ideal for investigating the covering factor and how it varies. In this paper we present an initial analysis of a small sample of 10 of these sources to investigate how well the covering factor can be determined from X-ray spectra. While this sample is small, it was selected in order to cover a wide range in \lx, paying particular attention to low (\lx$\sim10^{42}$ \ergs) and high (\lx$\sim10^{44}$ \ergs) luminosity sources to make it as representative as possible. In Section 2, we introduce X-ray spectral torus models and make comparisons between them. In Section 3, we describe the sample and the data analysis. In Section 4, we present our spectral fitting results. In Section 5, we discuss potential biases and systematics. In Section 6, we compare our results to previous results. In Section 7, we explore what physically influences the AGN covering factor. Finally, in Section 8, we present our conclusions. We assume a flat cosmological model with $H_{\rm 0}$=70 km s$^{-1}$ Mpc$^{-1}$ and $\Omega_{\Lambda}$=0.73. For measurement uncertainties on our spectral fit parameters we present the 90\% confidence limits given two interesting parameters ($\Delta\chi$=4.61).

\section{X-ray spectral torus models}
\label{sec_models}

We first briefly summarize the development of X-ray spectral torus models, detailing their similarities and differences. We then illustrate the differences in the X-ray spectra resulting from these models by fitting simulated spectra created from one model to other sets of public models and comparing the results.

\subsection{Model details}

In recent years a suite of new X-ray spectral models that describe the reprocessing of X-rays in a torus-shaped medium have been published \citep{ikeda09,murphy09,brightman11,liu14}. These models employ Monte-Carlo techniques to calculate spectra, simulating photoelectric absorption, fluorescence and Compton scattering and build on previous work by \cite{matt91}, \cite{leahy93}, \cite{ghisellini94}, \cite{nandra94monte} and \cite{yaqoob97}. The  aforementioned models differ mostly in the geometry of the torus considered and the treatment of the different components, be it direct, scattered or line emission. The underlying physics of the models is the same, however, with photoelectric cross-sections from \cite{verner96}, abundances from \cite{anders89} and fluorescent yields from \cite{bambynek72}. However, the physical geometry and treatment of Compton scattering differs somewhat. The latest edition from \cite{liu14} is the first to explicitly simulate a clumpy torus.

The model of \cite{ikeda09} considers a spherical torus, which consists of a spherical distribution of matter with a biconical void. The column density, \nh, through the torus is a function of the inclination angle, where the \nh\ is maximum when the torus is seen edge-on (90\degree\ inclination angle). This decreases to zero as the inclination angle decreases and reaches the opening angle, at which point the source can be seen unobscured. The opening angle is a free parameter with a range of 0-70\degree. The model separates into three components: the direct zeroth-order transmitted component,  the absorbed reflected component and the unabsorbed reflected component.

\cite{murphy09} consider a torus with a circular cross section and a fixed opening angle of 60\degree\ (covering factor of 0.5). This model, known as {\sc mytorus}, also has separate components for the direct, scattered and line emission, motivated so that time variability between these different emissions can be studied. The line components are separated so as to account for instrumental systematic effects on the line energies. Both {\sc mytorus} and \cite{ikeda09} consider \nh\ up to 10$^{25}$ \cmsq.  

A new implementation of {\sc mytorus} was introduced by \cite{yaqoob12} in which the various model components can be used independently in so called `decoupled' mode, giving it added flexibility. In this way, it can be used to model a clumpy distribution of matter with an arbitrary covering factor.

The model of \cite{brightman11} also assumes a spherical torus with a biconical void, as in \cite{ikeda09}. However, the line of sight \nh\ through the torus is constant and not dependent on the inclination angle. This model also allows for a variable opening angle, ranging from 26--84\degree, and has the added advantage of extending up to \nh=10$^{26}$ \cmsq, thus allowing investigations of the most extreme obscuration. We refer to this model simply as the {\sc torus} model. In addition to this {\sc torus} model, \cite{brightman11} present a model for the special case where the source is fully covered (a spherical geometry) and includes variable elemental and iron abundances with respect to hydrogen. We refer to this model as the {\sc sphere} model.

Recently \cite{liu14} have investigated the effect of a clumpy medium, distributed in a toroidal geometry, on the emergent X-ray spectra. The parameters they consider are the volume filling factor of the clumps, the number of clumps along the line of sight, and the line of sight \nh. They also only consider \nh\ up to 10$^{25}$ \cmsq. 

\begin{figure*}
\begin{center}
\includegraphics[width=180mm]{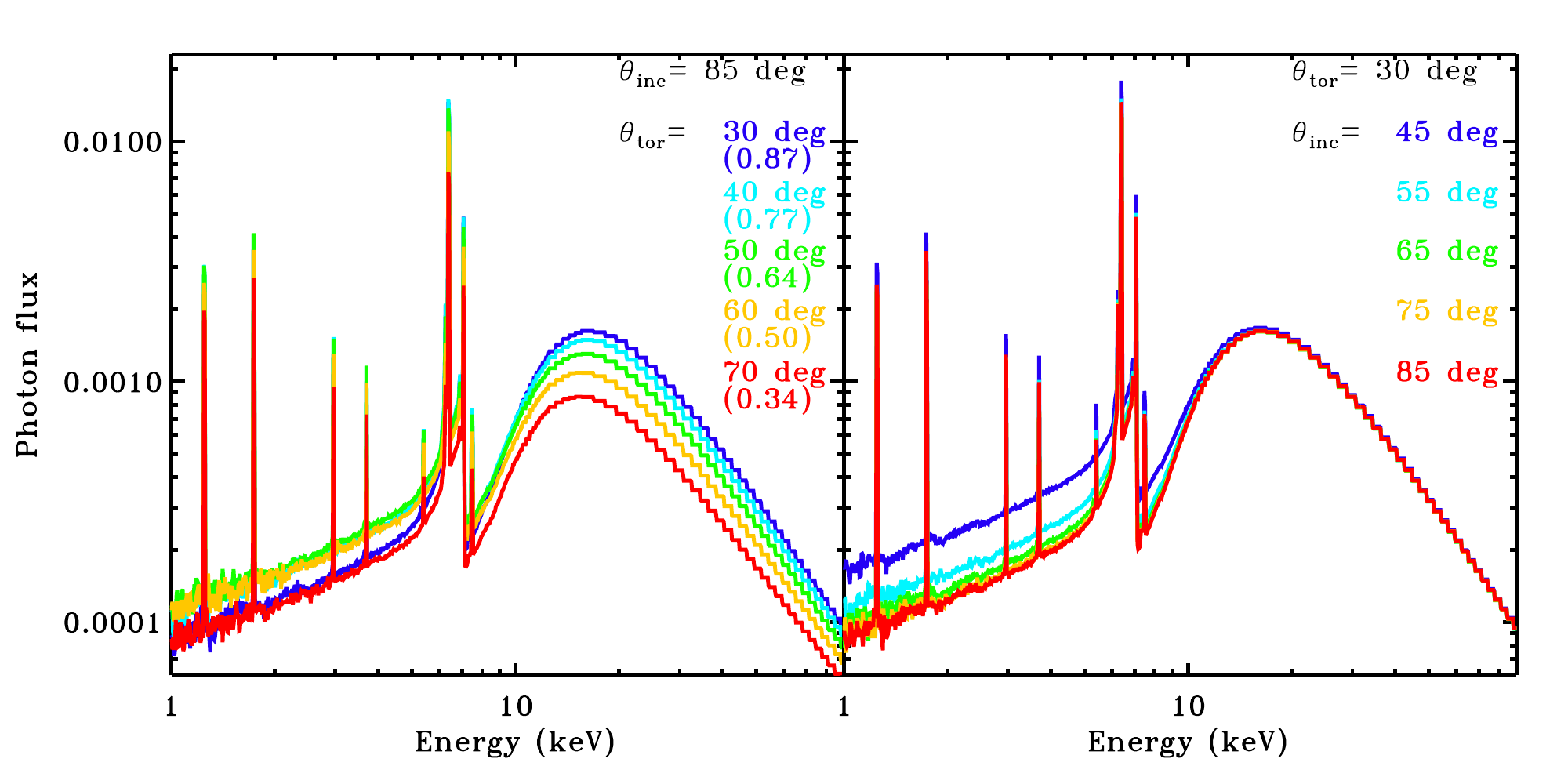}
\caption{X-ray {\sc torus} spectra for \nh=$2\times10^{24}$ \cmsq, $\Gamma$=2 and different torus opening angles, \thetator\ (left), where the equivalent covering factor is in parenthesis, and different inclination angles, $\theta_{\rm inc}$ (right). The variation in spectral shape with \thetator\ above 10 keV is clear, illustrating the capability of \nustar\ to determine the torus opening angle for CT AGN. For a given torus opening angle, changes in the inclination angle primarily affect the $<10$ keV X-ray spectrum.}
\label{fig_tormod}
\end{center}
\end{figure*}

In our investigation of the covering factor in CT AGN, we use the \cite{brightman11} {\sc torus} model because it includes a variable torus opening angle and extends to a higher \nh\ range. Henceforth we will refer to the half-opening angle of the torus, measured from the polar axis to the edge of the torus, as \thetator. The covering factor of the torus is simply calculated from \thetator\ following \fcov=cos(\thetator).  Figure \ref{fig_tormod} shows spectra from this model, where the left panel shows a range of \thetator\ values (equivalent to a range of \fcov\ values), and the right panel shows a range in inclination angles ($\theta_{\rm inc}$). The effect of the variation of the opening angle can clearly be seen in the left panel over the entire 3-79 keV spectral range of \nustar, in particular in the strength of the Compton hump at $\sim20$ keV. This illustrates how \nustar\ is well suited for studying the covering factor in CT AGN.

\subsection{Model comparisons}

As these X-ray torus models are still fairly novel, interpreting their results when applied to real data is not yet fully tested or understood. In order to interpret our results using the {\sc torus} model and to make inferences from them, we aim to better discern how this model relates to other commonly used models. We therefore compare the {\sc torus} model to the {\sc pexrav} model of \cite{magdziarz95}, which describes the Compton reflection of X-rays off a semi-infinite slab of cold material. The {\sc pexrav} model has been extensively used in the literature for describing the effect of Compton scattering/reflection in AGN spectra and for fitting the spectra of reflection-dominated, CT AGN. Furthermore, we compare the {\sc torus} model to {\sc mytorus}, described above. These two models have differing geometries and thus a comparison highlights the effects of geometry on the X-ray spectra. {\sc mytorus} has been used frequently in the analysis of \nustar\ spectra of heavily obscured AGN \citep[e.g.][]{puccetti14, gandhi14, arevalo14, balokovic14,bauer14}, and thus a comparison is valuable. We do not make a comparison to the \cite{ikeda09} model or the clumpy torus models of \cite{liu14} as their models are not public. 

To make the comparisons, we simulate \nustar\ spectra from the {\sc torus} model for \nh=10$^{24}$, 10$^{25}$ and 10$^{26}$ \cmsq\ and for \thetator=40\degree, 60\degree\ and 80\degree\ for $\Gamma=1.9$ and a 3-79 keV flux of $\sim10^{-11}$ \ergcms. We simulate \nustar\ spectra with 100 ks exposures without counting statistics in order to only consider the effect of the model parameters on the spectrum, and we use response and background files from a randomly selected observation, that of NGC~1320 (see Table \ref{table_obsdat}, obsid 60061036004). As a consistency check, we fit the {\sc torus} model to the simulated data, where we find that the input parameters are recovered.

\begin{figure*}
\begin{center}
\includegraphics[width=160mm]{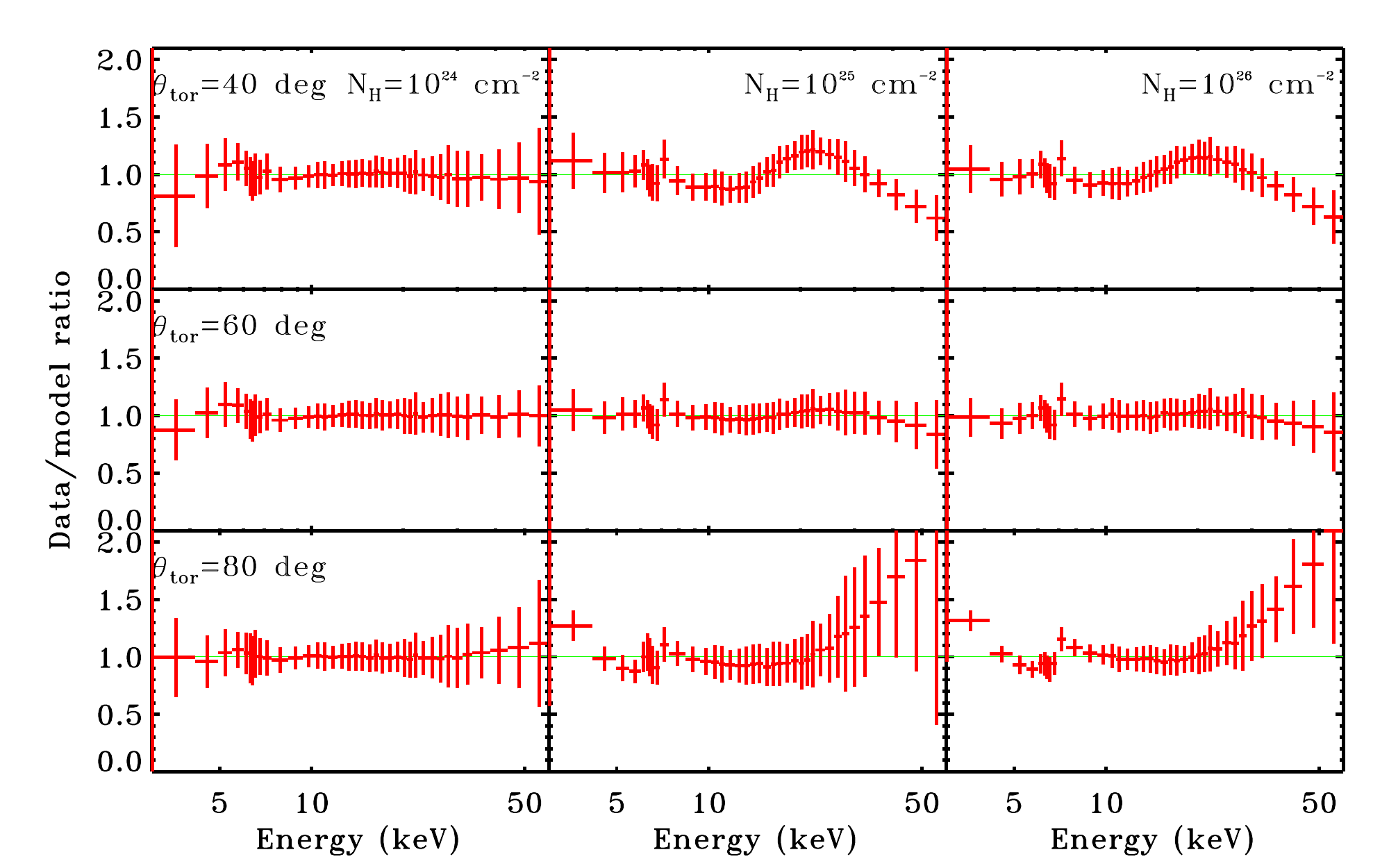}
\label{fig_simspec}

\caption{Data-to-model ratios of simulated \nustar\ spectra using the {\sc torus} model (red data points) fitted by {\sc pexrav} plus an absorbed power-law for \nh=10$^{24}$ \cmsq\ , for three {\sc torus} \nh\ values (\nh=10$^{24}$, 10$^{25}$ and 10$^{26}$ \cmsq) and three values of \thetator\ (40\degree, 60\degree\ and 80\degree). For {\sc torus} \nh$=10^{24}$ \cmsq, where the optical depth to Compton scattering is just below unity, the fitted models agree with the {\sc torus} model. This is also the case for all \nh\ values with a 60\degree\ opening angle. However at small (40\degree) and large (80\degree) opening angles, {\sc pexrav} cannot reproduce the shape of the {\sc torus} model. }

\includegraphics[width=160mm]{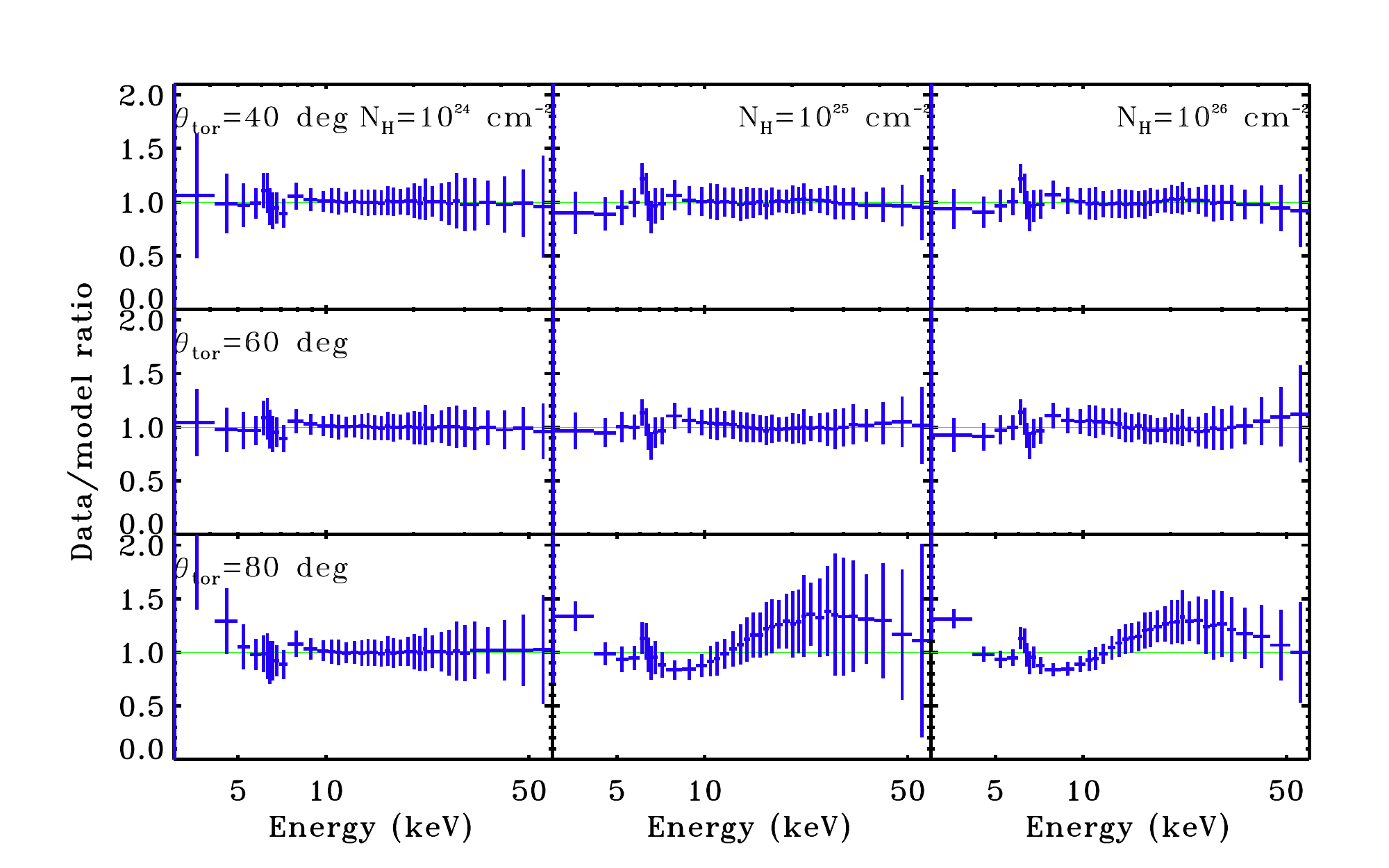}
\caption{Same as Fig. \ref{fig_simspec}, but simulated spectra fitted by {\sc mytorus} in decoupled mode.}

\label{fig_simspec2}
\end{center}
\end{figure*}

We subsequently fit each simulated spectrum between 3-79 keV with the pure reflection component of {\sc pexrav} with the high-energy cut off in the continuum set to the maximum ($1\times10^{6}$ eV), the abundances at solar and the cosine of the inclination angle at 0.45. We add a narrow Gaussian component at 6.4 keV to model the Fe K$\alpha$ line, since unlike {\sc torus}, {\sc pexrav} does not self-consistently include Fe fluorescence. For \nh=10$^{24}$ \cmsq, an additional absorbed power-law component is required, which we model with {\sc zwabs $\times$ powerlaw}, the photon index of which is fixed to that of {\sc pexrav}. Figure \ref{fig_simspec} shows how the best fit model compares to the simulated spectra. For {\sc torus} \nh$=10^{24}$ \cmsq, where the optical depth to Compton scattering is just below unity, the fitted models agree with the {\sc torus} model, with no obvious deviations from unity in the data-to-model ratio. This is also the case for all \nh\ values with a 60\degree\ opening angle. However, at small (40\degree) and large (80\degree) opening angles, when fitting with {\sc pexrav} there are obvious deviations in the data-to-model ratio above 10 keV. This implies that {\sc pexrav} is not able to reproduce the varying high-energy spectral shapes produced by a torus geometry. 

Following this we fit {\sc mytorus} to these simulated spectra in both coupled and decoupled mode. In coupled mode the parameters of the scattered, direct and line components are fixed to one another and thus the covering factor is 0.5. For the decoupled mode we combine the direct absorbed component with two scattered and line components, one where the inclination is fixed to 0\degree, the other where it is fixed to 90\degree, with all parameters linked with the exemption of the normalisations. This implementation represents backwards scattering and forwards scattering respectively, which as described by \cite{yaqoob12}, mimics an absorber with a clumpy distribution of matter with a variable covering factor. The data-to-model ratios for the fit with {\sc mytorus} in decoupled mode are shown in Fig. \ref{fig_simspec2}. These show that the flexibility of {\sc mytorus} in this mode can reproduce the spectral shape of the {\sc torus} model for a wide range of parameters, showing deviations only for the extreme case of \thetator=80\degree\ and \nh=10$^{25}$ \cmsq\ and above.

\begin{figure}
\begin{center}
\includegraphics[width=90mm]{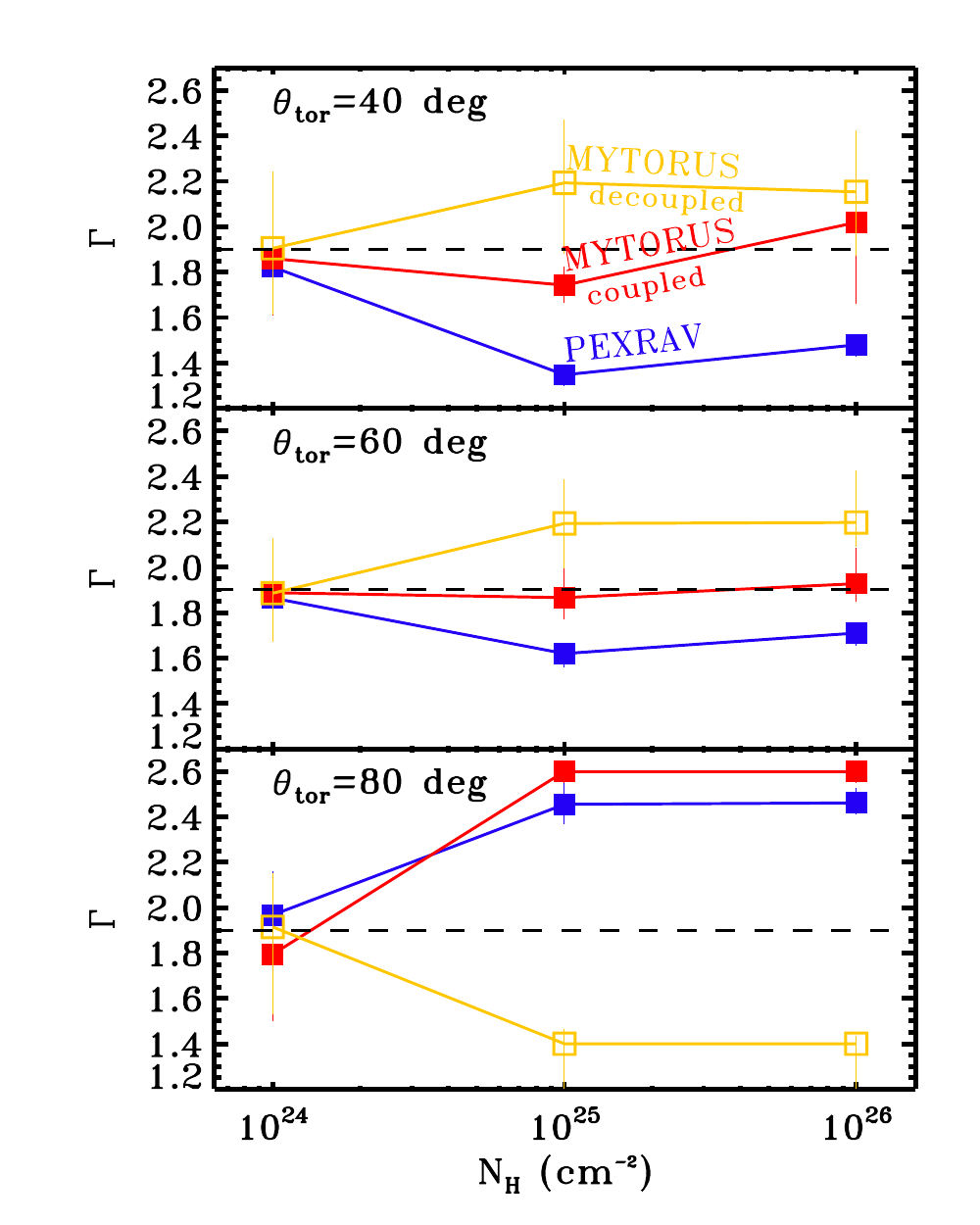}
\caption{Comparison of the power-law index, $\Gamma$, determined from fitting {\sc pexrav} and {\sc mytorus} to simulated \nustar\ spectra using the {\sc torus} model ($\Gamma$=1.9, dashed line) for \nh=10$^{24}$, 10$^{25}$ and 10$^{26}$ \cmsq\ and three opening angles (40\degree, 60\degree\ and 80\degree). }
\label{fig_simgam}
\end{center}
\end{figure}

Since the power-law index, $\Gamma$, is a good indicator of the overall shape of the spectrum, we compare that parameter for the fitted models to the value of $\Gamma$ adopted by the input model. We do not aim to test if the fitted models can recover the intrinsic $\Gamma$, since the models have different geometries, rather we explore the differences in $\Gamma$ in order to gain insight into the various spectral shapes produced by each model. This will also allow better interpretation of the parameters obtained when fitting these models to real data. Figure \ref{fig_simgam} shows how the best-fit $\Gamma$ from {\sc pexrav} and {\sc mytorus} compare to the input value of 1.9 from the {\sc torus} model given the input \nh\ and \thetator\ parameters. For all models, the measured $\Gamma$ is consistent with the input $\Gamma$ at \nh=$10^{24}$ \cmsq. For {\sc pexrav}, this diverges to lower $\Gamma$ values at higher \nh\ for 40\degree\ and 60\degree\ opening angles, indicating that the {\sc torus} model is producing a stronger Compton hump than {\sc pexrav} for these parameters. {\sc mytorus} in coupled mode is roughly consistent at 40\degree, whereas 60\degree\ shows no difference. This is expected since the opening angle in this model is indeed 60\degree. For 80\degree, both {\sc pexrav} and {\sc mytorus} measure rather larger $\Gamma$ values, indicating that the Compton hump is weaker in the {\sc torus} model for large \thetator. 

For {\sc mytorus} in decoupled mode, as in coupled mode, the measured $\Gamma$ from the simulated spectra agree with the input value within the measurement uncertainties for the case of 40\degree\ and 60\degree. The measurement uncertainties are higher in decoupled mode due to the larger number of degrees of freedom with this implementation. For the case of 80\degree\ there is a large discrepancy between the input and recovered $\Gamma$ values.

As described above, the decoupled implementation of {\sc mytorus} allows for an arbitrary covering factor. Following \cite{yaqoob12}, \cite{puccetti14} estimated the covering factor for NGC~4945 from \fcov$\sim0.5\times A_{\rm S90}/A_{\rm Z90}$, the ratio of the normalizations of the direct and scattered components at 90\degree\ inclination angles. We can directly compare this estimation with our input covering factors. Our input \thetator\ values correspond to \fcov=0.77, 0.50 and 0.17. From the above formulation, we find the covering factors estimated by the decoupled implementation of {\sc mytorus} to be 0.41, 0.09 and 0 respectively, which are much lower than the input values. It does however recover the input trend of decreasing covering factors. The decoupled implementation of {\sc mytorus} only mimics a free covering factor and thus an agreement was not necessarily expected.

Along with the {\sc pexrav} deviations in spectral shape shown in Figure \ref{fig_simspec}, the above analysis has shown that {\sc pexrav} will also systematically obtain different spectral parameters compared to the {\sc torus} model. We have also shown that the opening angle is an important parameter in determining the spectral shape, and thus models with fixed opening angles, such as {\sc mytorus}, will also systematically obtain different parameters. We therefore conclude that slab reflection models should not be used for fitting the high-energy X-ray spectra of CTAGN, as concluded by \cite{murphy09}, and that ideally spectral models with the covering factor of the Compton-thick gas as a free parameter should be used.

\section{Sample properties and data analysis}

Our sample consists of 10 local ($z<0.1$) CT AGN observed by \nustar: NGC~424, NGC~1068, 2MFGC~2280, NGC~1320, NGC~1386, NGC~3079, IC~2560, Mrk 34, NGC~4945 and the Circinus galaxy. With the exception of 2MFGC~2280, these sources were selected to be known CT AGN and to cover a wide range in X-ray luminosity. Particular attention was paid to selecting sources in the low (\lx$\sim10^{42}$ \ergs) and high ($\sim10^{44}$ \ergs) luminosity regimes, thus enveloping the bulk of the local CT AGN population, which is at moderate luminosities, so that our sample could be as representative as possible, despite its small size. CT AGN are selected since the obscuration is required to be optically thick to Compton scattering in order for the models to be effective at distinguishing the covering factor. The models are degenerate with covering factor for Compton-thin media. Previously published \nustar\ observations were included in order to facilitate comparison among models. 

Observations of NGC~424, NGC~1320 and IC~2560 are presented in \cite{balokovic14} with modelling using {\sc pexrav} and {\sc mytorus}. Results on NGC~1068 are presented in \cite{bauer14}, also with modelling using {\sc pexrav} and {\sc mytorus}. Mrk 34 is presented in \cite{gandhi14} with modelling using {\sc mytorus} and {\sc torus}. NGC~4945 is presented in \cite{puccetti14} and Circinus is presented in \cite{arevalo14}; both are modelled using {\sc mytorus}. The \nustar\ observations of NGC~1386, NGC~3079 and 2MFGC~2280 are presented here for the first time. No detailed X-ray spectral modelling has previously been published on 2MFGC~2280, and it was selected for study here based on its flat spectral shape in the \nustar\ band, which resembles that of the other obscured AGN considered in this work. 

Table \ref{table_obsdat} summarizes the basic observational data for this sample. The \nustar\ observations were performed for several different scientific purposes \citep[see][for a description of the \nustar\ survey programs]{harrison13} and therefore differ significantly in exposure time and signal-to-noise ratio. We use the first three existing \nustar\ observations of NGC~1068, both observations of NGC~1320 and both observations of IC~2560 as these sources are not known to be variable in the hard X-rays. NGC~4945 is known to be variable. We mitigate the effect of the variability by analysing only one \nustar\ obsID from this source, of which we choose the first of three. We investigate the effect on our conclusions of this choice later in the paper. For a time resolved-analysis of this source, we refer the reader to \cite{puccetti14}.

For NGC~424, NGC~1320, NGC~1386, NGC~3079, IC~2560 and Mrk 34, we use additional archival data from \xmm\ from 0.5-10 keV to provide additional spectral constraints. \xmm\ data were chosen specifically over other instruments as \xmm\ and \nustar\ are well matched in terms of effective area and spatial resolution in the energy range where they commonly observe (i.e. 3-10 keV). No good quality data exist for 2MFGC~2280 below 10 keV. NGC~1068, NGC~4945 and Circinus are part of a different, non-snapshot \nustar\ program and hence have longer exposure times and higher signal-to-noise spectra and are thus of sufficient quality to constrain the torus parameters well without additional soft X-ray data.

\begin{table*}
\centering
\caption{Summary of the observational data used in this analysis.}
\label{table_obsdat}
\begin{center}
\begin{tabular}{l r r l r r r r}
\hline
Source name 	& RA & Dec & telescope & obsid & date & exposure & net count rate \\
			& (deg) & (deg)	&		&	&	& (ks)	& (counts s$^{-1}$) \\
(1) & (2) & (3) & (4) & (5) & (6) & (7) & (8) \\
\hline

NGC 424 		&  17.86511 	& -38.08347 	& \nustar\ 	& 60061007002	& 2013-01-26	& 15.5 	& 0.052/0.053 \\
			&			&			& \xmm\	& 2942301		& 2001-12-10	& 4.5		& 0.236 \\
NGC 1068 	&  40.66963 	&  -0.01328 	& \nustar\	& 60002030002 	& 2012-12-18	& 57.8	& 0.112/0.107 \\ 						& 			& 			&		& 60002030004 	& 2012-12-20	& 48.6	& 0.111/0.111 \\
			& 			& 			&		& 60002030006 	& 2012-12-21	& 19.5	& 0.115/0.113 \\
2MFGC 2280 	&  42.67746 	&  54.70489 	& \nustar\	& 60061030002 	& 2013-02-16	& 15.9	& 0.019/0.018 \\
NGC 1320 	&  51.20292 	&  -3.04228 	& \nustar\ & 60061036002 	& 2012-10-25	& 14.5	& 0.028/0.023 \\
		 	&  		 	&  		 	& 		& 60061036004 	& 2013-02-10	& 28.0	& 0.025/0.022 \\
			&			&			& \xmm\	& 405240201		& 2006-08-06	& 12.5	& 0.081 \\	 
NGC 1386 	&  54.19242 	& -35.99941 	& \nustar\	& 60001063002 	& 2013-07-19 	& 21.2	& 0.010/0.011 \\
			&			&			& \xmm\	& 140950201		& 2002-12-29	& 13.9	& 0.153 \\
NGC 3079	& 150.49085	& 55.67979	& \nustar\	& 60061097002	& 2013-11-12	& 21.5	& 0.062/0.063 \\
			&			&			& \xmm\	& 110930201		& 2001-04-13	& 13.6	& 0.092 \\  		
IC 2560 		& 154.07799 	& -33.56379 	& \nustar\	& 50001039002 	& 2013-01-28	& 23.4	& 0.012/0.012 \\
			&			&			& 		& 50001039004	& 2014-07-16	& 49.6	& 0.012/0.012 \\
			&			&			& \xmm\	& 0203890101		& 2003-12-26	& 80.5	& 0.006/0.007 \\
Mrk 34 		& 158.53580 	&  60.03111 	& \nustar\	& 60001134002 	& 2013-09-19	& 23.9	& 0.007/0.007 \\
			&			&			& \xmm\	& 0306050701		& 2005-04-04	& 8.8/22.9	& 0.070/0.013 \\
NGC 4945 	& 196.36449 	& -49.46821 	& \nustar\	& 60002051002 	& 2013-02-10	& 45.2	& 0.263/0.249 \\
Circinus		& 213.29146	& -65.33922	& \nustar\ & 60002039002	& 2013-01-25	& 53.8	& 1.046/0.984 \\

\hline
\end{tabular}
\tablecomments{Column (1) gives the source name, Columns (2) and (3) list the J2000 position given in NED in degrees, column (4) gives the telescope name, column (5) lists the observation ID, column (6) gives the start date of the observation, column (7) gives the exposure time in ks and column (8) gives the net count rate for each instrument, be it FPMA/B for \nustar\ (3-79 keV) or pn/MOS for \xmm\ (0.5-10 keV). Only pn data are used for NGC~424, NGC~1320, NGC~1386 and NGC~3079 and only MOS data are used for IC~2560. }
\end{center}
\end{table*}

The raw data were reduced using the {\sc NuSTARDAS} software package version 1.2.1. The events were cleaned and filtered using the {\tt nupipeline} script with standard parameters\footnote{The NuSTAR Data Analysis Software Guide, http://heasarc.gsfc.nasa.gov/docs/nustar/analysis/\\NuSTARDAS\_swguide\_v1.3.pdf}. The {\tt nuproducts} task was used to generate the spectra and the corresponding response files. Spectra were extracted from circular apertures centred on the peak of the point source emission, with radii between 30\arcsec and 90\arcsec (larger radii for higher signal-to-background ratio). The background spectra were extracted from regions encompassing the same detector as the source, excluding the source extraction region and avoiding the wings of the PSF as much as possible. Data from both focal plane modules (FPMA and FPMB) and from multiple observations are used for simultaneous fitting, without co-adding. Additional details of the data reduction for specific sources can be found in the papers listed above. The observations of NGC~1386 and 2MFGC~2280, which are not published elsewhere, are reduced in a fashion similar to \cite{balokovic14}.

The \xmm\ data reduction for the observations of NGC~424, NGC~1320, NGC~1386 and NGC~3079 were described in \cite{brightman11}, where only EPIC-pn data were used, with source events extracted from 35\arcsec\ radius circular regions. The events were then filtered for background flares when the level of the background count rate was determined to be twice the level at which the excess variance was determined to be zero. The \xmm\ data reduction for the observation of Mrk 34 is described in \cite{gandhi14} and the \xmm\ data reduction for IC~2560 is described in \cite{tilak08}.

All spectra were grouped with a minimum of 20 counts per bin using the {\sc heasarc} tool {\tt grppha}. We use {\sc xspec} version 12.6.0 to carry out X-ray spectral fitting, and the $\chi^2$ statistic as the fit statistic, with the background subtracted. For \nustar\ data, only energies from 3-79 keV were considered, as the calibration at lower energies is uncertain, and the \nustar\ response cuts off at 79 keV due to an absorption edge in the optics. \xmm\ data were used from 1-10 keV, allowing considerable overlap with \nustar. 

The primary goal of our analysis is to determine the covering factor of the obscuring material in these CT AGN using the {\sc torus} model. The AGN emission which is described by this model is by far the dominant emission in the \nustar\ band for these sources, thus all spectra were fitted with the {\sc torus} model as a baseline. We fix the inclination angle of the torus to the edge-on position of 87\degree, the maximum inclination angle allowed by the model. As described above, the X-ray spectrum above 10 keV is largely insensitive to the inclination angle, when the inclination angle is greater than \thetator. Fixing the inclination angle thus reduces the number of free parameters in the fit and the edge-on choice allows the exploration of the full range of opening angles as opening angles can only be determined up to the inclination angle, after which, the source becomes unobscured. \cite{arevalo14} find that when fitting the {\sc torus} model to the spectrum of Circinus, that both \thetator\ and the inclination angle can be constrained and provide a better fit to the data than if the inclination angle were fixed. We thus allow the inclination angle to be free when fitting the spectrum of Circinus.

While the torus emission is the dominant component in the \nustar\ band, emission not directly associated with the intrinsic AGN emission or the reprocessing in the torus, such as soft emission from photo-ionised material, radiative recombination \citep{guainazzi07} and Thompson-scattered AGN light, are common in obscured AGN and are non-negligible even in the \nustar\ band. Furthermore, due to the size of \nustar's PSF, X-ray sources in the host galaxy of the AGN may also contribute. For example, \cite{puccetti14} finds that $\sim60$\% of emission in the $4-6$ keV band comes from extra-nuclear sources within the \nustar\ extraction region, while \cite{bauer14} finds that 28\% of the Fe K$\alpha$ emission from NGC 1068 comes from the host galaxy. It is expected, however, that these extra-nuclear sources contribute far less at higher energies. We therefore add a power-law model and a thermal plasma component modelled by {\sc apec} to broadly account for these soft excess components and extra-nuclear sources. We allow the secondary power-law index to vary freely, likewise for the temperature of the {\sc apec} model. For 2MFGC~2280 where only \nustar\ data are available, the soft components were not statistically required, so they were removed from the fit. No secondary power-law was required in the fit for IC~2560. For NGC~1068, NGC~4945 and Circinus where only \nustar\ data were used, the {\sc apec} model was not required. This is expected since this model generally describes emission outside the \nustar\ bandpass. The spectra of NGC~424 and NGC~1320 also did not require the {\sc apec} model.

NGC~1068 has a well known Fe complex at 6-7 keV, consisting of neutral Fe K$\alpha$ and $\beta$ emission, plus ionised emission at 6.7 and 6.96 keV. While the {\sc torus} model accounts for the neutral emission, we add Gaussian components fixed at the energies of the ionised emission to account for these lines, where the widths are fixed at small values (1 eV). As found by \cite{sambruna11} and \cite{molendi03} for Circinus and \cite{yaqoob12} for NGC~4945, we also find that the \nustar\ spectra of these AGN require an additional Gaussian component each to model lines in the Fe complex, fixed at 6.7 keV in NGC~4945 and allowed to vary around 6.6 keV for Circinus, as required by the data. Additional Gaussian components at the energies of the Fe complex were required to fit the spectra of NGC~1320, NGC~1386 and IC~2560. We summarize all the models used for each source in Table \ref{table_cnomo}.

We allow the cross-normalization between the two \nustar\ focal plane modules (FPMs) to vary in the fitting to account for instrumental cross-calibration. We also allow the cross-normalization between the FPMs and \xmm\ to vary to account for instrumental cross-calibration and the fact that the the \xmm\ observations are not simultaneous with \nustar. We fix the other parameters to each other since there is a known agreement between spectral parameters determined from \xmm\ and \nustar\ \citep[e.g.][]{walton13,walton14}. We present the cross-normalizations in Table \ref{table_cnomo}. The observations are mainly consistent with the cross-normalization between the FPMs being unity, with the exceptions being NGC~1068, NGC~1320 and Circinus. NGC~1068 and Circinus both show that the normalization of  FPMB to be 3\% greater than the normalization of the FPMA. For NGC~1320 the FPMB normalization is 9\% lower. \cite{balokovic14} also investigate the cross normalization between the two FPMs for NGC~424, NGC~1320 and IC~2560, and our results agree. Full details of \nustar\ in-orbit calibration is presented in Madsen et al. (in preparation).

\begin{table*}
\centering
\caption{Details of the cross-normalizations.} 
\label{table_cnomo}
\begin{center}
\begin{tabular}{l l l l}
\hline
Source name 	& FPMB/FPMA & \xmm/FPMA & Models used \\
(1) & (2) & (3) & (4) \\
\hline

NGC 424		& 1.07$_{-0.12}^{+0.13}$ & 0.97$_{-0.15}^{+0.16}$	& {\sc torus+powerlaw} \\
NGC 1068 	& 1.03$_{-0.02}^{+0.03}$ & - 	& {\sc torus+powerlaw+zgauss+zgauss} \\
2MFGC2280	& 1.02$_{-0.14}^{+0.16}$ & - 	& {\sc torus} \\
NGC 1320	& 0.91$_{-0.07}^{+0.07}$ & 0.89$_{-0.12}^{+0.13}$	& {\sc torus+powerlaw+zgauss} \\
NGC 1386	& 0.80$_{-0.15}^{+0.20}$ & 0.84$_{-0.16}^{+0.17}$ & {\sc torus+powerlaw+apec+zgauss} \\
NGC 3079	& 1.06$_{-0.10}^{+0.11}$ & 0.65$_{-0.14}^{+0.16}$ & {\sc torus+powerlaw+apec} \\
IC 2560		& 1.07$_{-0.09}^{+0.09}$ & 0.82$_{-0.11}^{+0.16}$, 0.92$_{-0.12}^{+0.17}$ & {\sc torus+apec+zgauss} \\
Mrk 34		& 1.09$_{-0.22}^{+0.27}$ & 0.93$_{-0.27}^{+0.37}$, 0.81$_{-0.24}^{+0.37}$ & {\sc torus+powerlaw+apec} \\
NGC 4945	& 1.02$_{-0.02}^{+0.02}$ & - & {\sc torus+powerlaw+zgauss}  \\
Circinus		& 1.03$_{-0.01}^{+0.01}$ & - & {\sc torus+powerlaw+zgauss} \\

\hline
\end{tabular}
\tablecomments{Column (2) gives the cross-normalization between the two FPMs, column (3) gives the cross-normalization between FMPA and \xmm\ (MOS 1 and MOS 2 respectively for IC~2560 and pn and combined MOS for Mrk 34) and column (4) lists the models used to fit each of the spectra.}
\end{center}
\end{table*}

\section{Spectral fitting results}

Our spectral fitting with the {\sc torus} model reproduces the data well in all ten sources, with reasonable reduced \chisq. We confirm the Compton-thick nature of all 10 AGN, with \nh\ constraints in excess of $1.5\times10^{24}$ \cmsq. The Compton-thick nature of 2MFGC~2280 is shown here for the first time. 

The best-fit spectral parameters are presented in Table \ref{table_specpar}, along with their 90\% confidence intervals calculated using a $\Delta\chi^2=4.61$ criterion on two interesting parameters in order to minimize degeneracies. The uncertainty on the luminosity is propagated from the uncertainty on the normalization of the {\sc torus} model. Details of the parameters of the components used to fit the soft X-ray excesses are listed in Table \ref{table_specpar3}.

The best fit \thetator\ measured in our sample spans a large range, from 26--80\degree, limited by the allowed range of the model. The uncertainties on \thetator\ range from 5--40\degree, with the best constraints coming from NGC~1068, NGC~4945 and Circinus which have the highest signal-to-noise spectra in the sample. While the fit to Mrk 34 gives a fairly large \thetator, the constraints are poor, with statistically allowed values ranging from 26--80\degree. The case is similar for 2MFGC~2280, albeit with slightly tighter constraints. 

\begin{figure}
\begin{center}
\includegraphics[width=90mm]{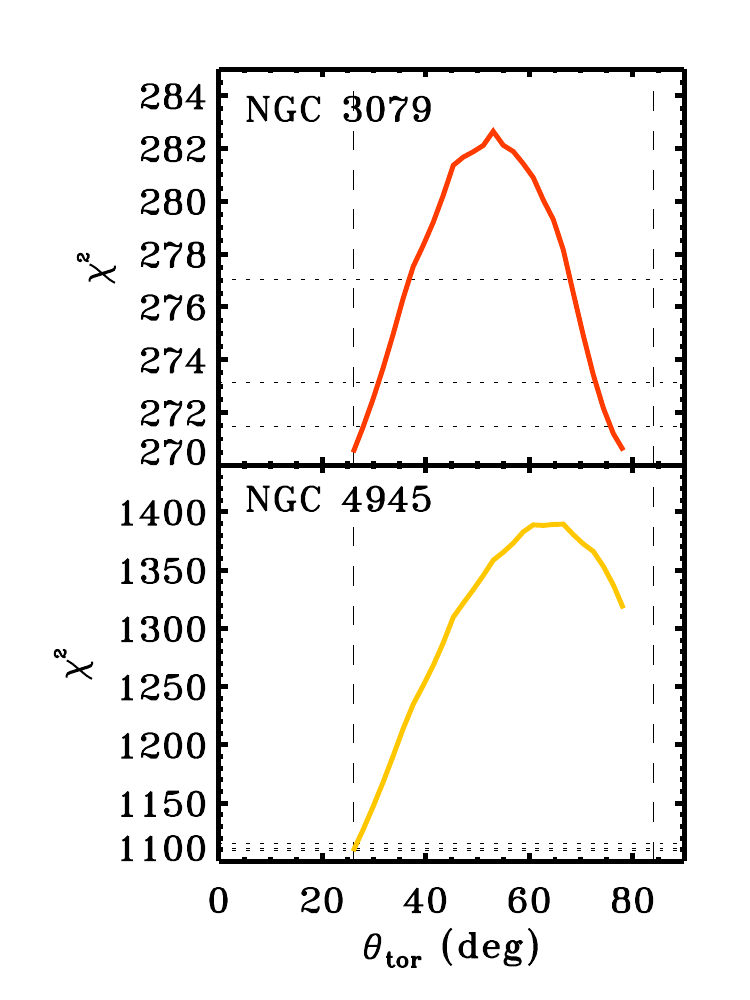}
\caption{\chisq\ as a function of \thetator\ for fits to NGC~3079 and NGC~4945, showing local minima at both small and large opening angles. For NGC~3079 the large opening angle is marginally favoured, while in NGC~4945 the small opening angle is significantly favoured. The dashed lines indicate the minimum (26\degree) and maximum (84\degree) \thetator\ values allowed by the model and the dotted lines show the 68, 90 and 99\% confidence levels which correspond to $\Delta\chi^2$ values of 1.00, 2.71 and 6.63 respectively.}
\label{fig_thetacon}
\end{center}
\end{figure}

In two sources, NGC~3079 and NGC~4945, we discover bimodality in statistic space for \thetator, with local minima at small and large opening angles. We show their \chisq\ as a function of \thetator\ in Figure \ref{fig_thetacon}. For NGC~3079, the large opening angle provides a marginally better fit, whereas for NGC~4945 the small opening angle is a significantly better fit.

For two sources, NGC~1386 and NGC~4945, we find that the best-fit \thetator\ reaches the lower limit on this parameter (26\degree), which implies that both the sources are highly covered. We also find that the $\Gamma$ values for these fits deviate significantly from the canonical value of 1.9 (steep at 2.52 for NGC~1386 and flat for NGC~4945 at 1.58). The fact that the deviations from the canonical value occur at the edge of the \thetator\ parameter space suggests that the true \thetator\ of these sources lies outside of the range of the {\sc torus} model (i.e. \thetator$<$26\degree).  

\cite{brightman11} also present a model of a fully covered source (i.e. 0\degree\ opening angle), which includes variable iron and elemental abundances, known as {\sc sphere}. We fit this model, with the iron abundance free, in place of the {\sc torus} model for NGC~1386 and NGC~4945, and present the results in Table \ref{table_specpar}. This is also done for NGC~3079 since the small opening angle also provides a good fit to that source. An improvement in the fit statistic is found for both NGC~3079 and NGC~4945, where the spherical model produces a significantly better ($>$90\% significance) fit than the {\sc torus} model for both ($\Delta\chi^2$=4 and 136, respectively). The iron abundance has been estimated to be slightly higher, albeit statistically consistent with solar metalicity. The $\Gamma$ inferred from the {\sc sphere} model is also consistent with the canonical value for both sources. For NGC~1386, the {\sc sphere} model produces extreme $\Gamma$ and iron abundances values and iherefore it is not likely to be a good description of the spectrum. 

Due to the range of \thetator\ allowed for the {\sc torus} model, we cannot test if a very large opening angle ($\sim$90\degree) would provide as equally good a solution for NGC~3079 or NGC~4945. However for the local minimum at the largest angle, the $\Gamma$ for NGC~4945 is even flatter (1.43) than the smallest opening angle and does not suggest that a large opening angle would describe the spectrum well. For this solution, \nh=2.64$\times10^{24}$ \cmsq\ and \lx=2.27$\times10^{42}$ \ergs.

We conclude from this that NGC~3079 and NGC~4945 are heavily buried sources with a covering factors close to unity. For all further analysis we assign a value of \thetator=0\degree\ to these sources. We discuss this result later, especially for NGC~4945 where previous results have favoured a very low covering factor \citep[e.g.][]{madejski00}.
 
\begin{table*}
\centering
\caption{Best-fit spectral parameters of the {\sc torus} and {\sc sphere} models.}
\label{table_specpar}
\begin{center}
\begin{tabular}{l l l l l l l l l l l}
\hline
\multicolumn{10}{c}{{\sc torus} fits} \\
Source name & redshift & PHA bins & $\chi^2$ & $\chi^2_r$ & \nh\ & $\Gamma$ & $\theta_{\rm tor}$ & log$_{10}$\fx\ & log$_{10}$\lx \\
(1) & (2) & (3) & (4) & (5) & (6) & (7) & (8) & (9) & (10) \\
\hline

NGC 424 & 0.0118 &  121 &  139.27 &    1.23 &   5.41$^{+15.67}_{- 1.23}$ &  2.20$^{+ 0.34}_{- 0.16}$ & 78.19$^{+ 0.43}_{-15.01}$ & -10.80 & 43.96$^{+ 0.21}_{- 0.16}$ \\
NGC 1068 & 0.0038 & 1140 & 1950.47 &    1.72 &   6.32$^{+ 0.44}_{- 0.05}$ &  2.31$^{+ 0.04}_{- 0.05}$ & 58.99$^{+ 4.51}_{- 2.76}$ & -10.51 & 42.87$^{+ 0.04}_{- 0.06}$ \\
2MFGC 2280 & 0.0150 &   36 &   13.18 &    0.43 &   2.50$^{+ 0.58}_{- 1.00}$ &  1.96$^{+ 0.26}_{- 0.56}$ & 78.97$^{+ 5.03}_{-41.87}$ & -10.98 & 43.26$^{+ 0.31}_{- 0.74}$ \\
NGC 1320 & 0.0089 &  172 &  190.37 &    1.18 & 100.00$^{+u}_{-70.81}$ &  1.66$^{+ 0.21}_{- 0.19}$ & 60.17$^{+10.69}_{-14.91}$ & -10.95 & 42.79$^{+ 0.12}_{- 0.09}$ \\
NGC 1386 & 0.0029 &   71 &   77.46 &    1.31 &   5.61$^{+ 2.11}_{- 1.16}$ &  2.92$^{+ 0.08}_{- 0.44}$ & 33.53$^{+10.79}_{- 7.23}$ & -11.68 & 41.84$^{+ 0.26}_{- 0.05}$ \\
NGC 3079 & 0.0037 &  214 &  270.38 &    1.33 &   2.37$^{+ 0.36}_{- 0.25}$ &  1.69$^{+ 0.11}_{- 0.15}$ & 79.75$^{+ 1.73}_{- 9.36}$ & -10.78 & 41.96$^{+ 0.19}_{- 0.33}$ \\
IC 2560 & 0.0096 &  165 &  182.51 &    1.19 & 100.00$^{+u}_{-86.64}$ &  2.53$^{+ 0.20}_{- 0.20}$ & 59.08$^{+ 9.57}_{-20.05}$ & -11.66 & 42.95$^{+ 0.11}_{- 0.13}$ \\
Mrk 34 & 0.0510 &   69 &   71.03 &    1.22 &  50.43$^{+49.57}_{-31.01}$ &  1.73$^{+ 1.25}_{- 0.56}$ & 72.51$^{+ 6.89}_{-46.51}$ & -11.52 & 44.18$^{+ 0.39}_{- 0.38}$ \\
NGC 4945 & 0.0019 &  906 & 1108.75 &    1.24 &   2.54$^{+ 0.13}_{- 0.15}$ &  1.58$^{+ 0.04}_{- 0.04}$ & 26.00$^{+ 0.50}_{-l}$ &  -9.65 & 41.92$^{+ 0.07}_{- 0.08}$ \\
Circinus & 0.0014 & 1715 & 1880.27 &    1.10 &   4.85$^{+ 0.39}_{- 0.42}$ &  2.27$^{+ 0.05}_{- 0.07}$ & 33.79$^{+ 1.83}_{- 1.56}$ &  -9.63 & 42.51$^{+ 0.07}_{- 0.09}$ \\

\hline
\multicolumn{10}{c}{{\sc sphere} fits} \\
Source name & redshift & PHA bins & $\chi^2$ & $\chi^2_r$ & \nh\ & $\Gamma$ & Fe abund. & log$_{10}$\fx\ & log$_{10}$\lx \\
(1) & (2) & (3) & (4) & (5) & (6) & (7) & (8) & (9) & (10) \\
\hline
NGC 1386 & 0.0029 &   71 &   73.74 &    1.25 &  0.76$^{+ 0.45}_{- 0.19}$ &  3.00$^{+u}_{-l}$ & 10.00$^{+u}_{- 5.44}$ & -11.56 & 39.67$^{+ 0.18}_{- 0.11}$ \\
NGC 3079 & 0.0037 &  214 &  266.42 &    1.31 &  1.84$^{+ 0.32}_{- 0.32}$ &  1.86$^{+ 0.30}_{- 0.23}$ &  1.42$^{+ 0.34}_{- 0.43}$ & -10.83 & 41.53$^{+ 0.45}_{- 0.43}$ \\
NGC 4945 & 0.0019 &  905 &  972.64 &    1.09 &  2.25$^{+ 0.24}_{- 0.07}$ &  1.78$^{+ 0.13}_{- 0.06}$ &  1.38$^{+ 0.16}_{- 0.37}$ &  -9.66 & 42.05$^{+ 0.17}_{- 0.10}$ \\

\hline
\end{tabular}
\tablecomments{Parameters determined by the {\sc torus} model are listed in the top rows. For NGC~1386, IC 3079 and NGC~4945 we also present the best-fit parameters from the {\sc sphere} model as the {\sc torus} model finds that these sources are highly covered (bottom rows). Column (1) lists the source name, column (2) gives the redshift of the source, column (3) shows the total number of pulse height analysis (PHA) bins used in the spectrum, column (4) gives the \chisq\ of the fit, column (5) gives the reduced \chisq, equal to \chisq\ divided by the number of degrees of freedom in the fit, column (6) gives the \nh\ in units of 10$^{24}$ \cmsq, with uncertainties, column (7) gives the photon-index, column (8) gives \thetator, determined by the {\sc torus} model in degrees or the iron abundance with respect to solar hydrogen abundance from the {\sc sphere} model, column (9) gives the logarithm of the observed 3-79 keV \nustar\ flux in \ergcms\ and column (10) gives the logarithm of the intrinsic (deabsorbed) rest-frame 2-10 keV luminosity in \ergs. $+u$ indicates that a parameter has reached the upper limit when estimating the uncertainty. $-l$ indicates that the lower limit has been reached.}
\end{center}
\end{table*}

\begin{table*}[H]
\centering
\caption{Best-fit spectral parameters of the secondary power-law, {\sc apec} and Gaussian line models.}
\label{table_specpar3}
\begin{center}
\begin{tabular}{l l l l l l l }
\hline
Source name & $\Gamma$ & A$_{\rm pl}/10^{-5}$	&  T$_{\rm apec}$ & A$_{\rm apec}/10^{-5}$ & E$_{\rm line}$ & A$_{\rm line}/10^{-5}$ \\
(1) & (2) & (3) & (4) & (5) & (6) & (7) \\
\hline
\multicolumn{7}{c}{{\sc torus} fits} \\

NGC 424 &  3.79$^{+ 3.45}_{- 0.93}$ &  11.27$^{+   4.00}_{-   5.78}$ & - &   - & - &   - \\
NGC 1068 &  2.29$^{+ 0.10}_{- 0.09}$ & 163.20$^{+  19.07}_{-  21.06}$ &  &   - &  &   - \\
2MFGC 2280 &   &   - &  &   - &  &   - \\
NGC 1320 &  3.70$^{+ 0.46}_{- 0.41}$ &  12.91$^{+   3.11}_{-   2.51}$ &  &   - & 6.55$^{+ 0.05}_{- 0.04}$ &   0.57$^{+   0.20}_{-   0.17}$ \\
NGC 1386 &  2.56$^{+ 1.72}_{- 2.76}$ &   1.61$^{+   1.07}_{-   1.06}$ & 0.68$^{+ 0.06}_{- 0.09}$ &   7.31$^{+   1.06}_{-   3.57}$ & 6.42$^{+ 0.04}_{- 0.06}$ &   0.47$^{+   0.17}_{-   0.17}$ \\
NGC 3079 &  1.95$^{+1.70}_{- 3.34}$ &   3.38$^{+   9.17}_{-   3.38}$ & 0.93$^{+ 0.36}_{- 0.10}$ &   9.92$^{+  21.45}_{-   5.72}$ &  &   - \\
IC 2560 &  - &   - & 0.68$^{+ 0.20}_{- 0.21}$ &   2.28$^{+   1.16}_{-   0.62}$ & 6.43$^{+ 0.02}_{- 0.01}$ &   0.67$^{+   0.16}_{-   0.15}$ \\
Mrk 34 &  2.90$^{+ 0.81}_{- 0.28}$ &   1.98$^{+   0.99}_{-   1.19}$ & 0.82$^{+ 0.38}_{- 0.05}$ &   0.92$^{+   0.49}_{-   0.54}$ & - &   - \\
NGC 4945 &  1.90$^{+ 0.55}_{- 0.48}$ &  29.72$^{+  38.76}_{-  16.28}$ & - &   - & - &   - \\
Circinus &  2.05$^{+ 1.96}_{- 1.19}$ & 226.28$^{+1698.37}_{- 226.28}$ & - &   - & 6.57$^{+ 0.04}_{- 0.00}$ &  12.86$^{+   1.17}_{-   1.12}$ \\

\hline

\multicolumn{7}{c}{{\sc sphere} fits} \\

NGC 1386 &  1.20$^{+ 0.33}_{- 0.30}$ &   1.78$^{+   0.85}_{-   0.63}$ & 0.68$^{+ 0.08}_{- 0.20}$ &   7.60$^{+   1.58}_{-   1.52}$ & 6.39$^{+ 0.05}_{- 0.04}$ &   0.54$^{+   0.22}_{-   0.23}$ \\
NGC 3079 &  1.38$^{+ 0.37}_{- 0.25}$ &   6.09$^{+   4.82}_{-   2.06}$ & 0.89$^{+ 0.33}_{- 0.61}$ &  11.82$^{+   2.55}_{-   5.60}$ &- &   - \\
NGC 4945 &  1.00$^{+ 0.22}_{- 0.25}$ &  12.57$^{+   5.80}_{-   4.41}$ & - &   - & - &   - \\

\hline

\end{tabular}
\tablecomments{Column (1) gives the source name, column (2) gives the power-law index of the secondary power-law, column (3) gives the normalization of the secondary power-law, column (4) gives the temperature of the {\sc apec} model in keV, column (5) gives the normalization of the {\sc apec} model, column (6) gives the energy of the Gaussian line in keV and column (7) gives the normalisation of the Gaussian line. }
\end{center}
\end{table*}

\begin{figure*}
\begin{center}
\includegraphics[width=180mm]{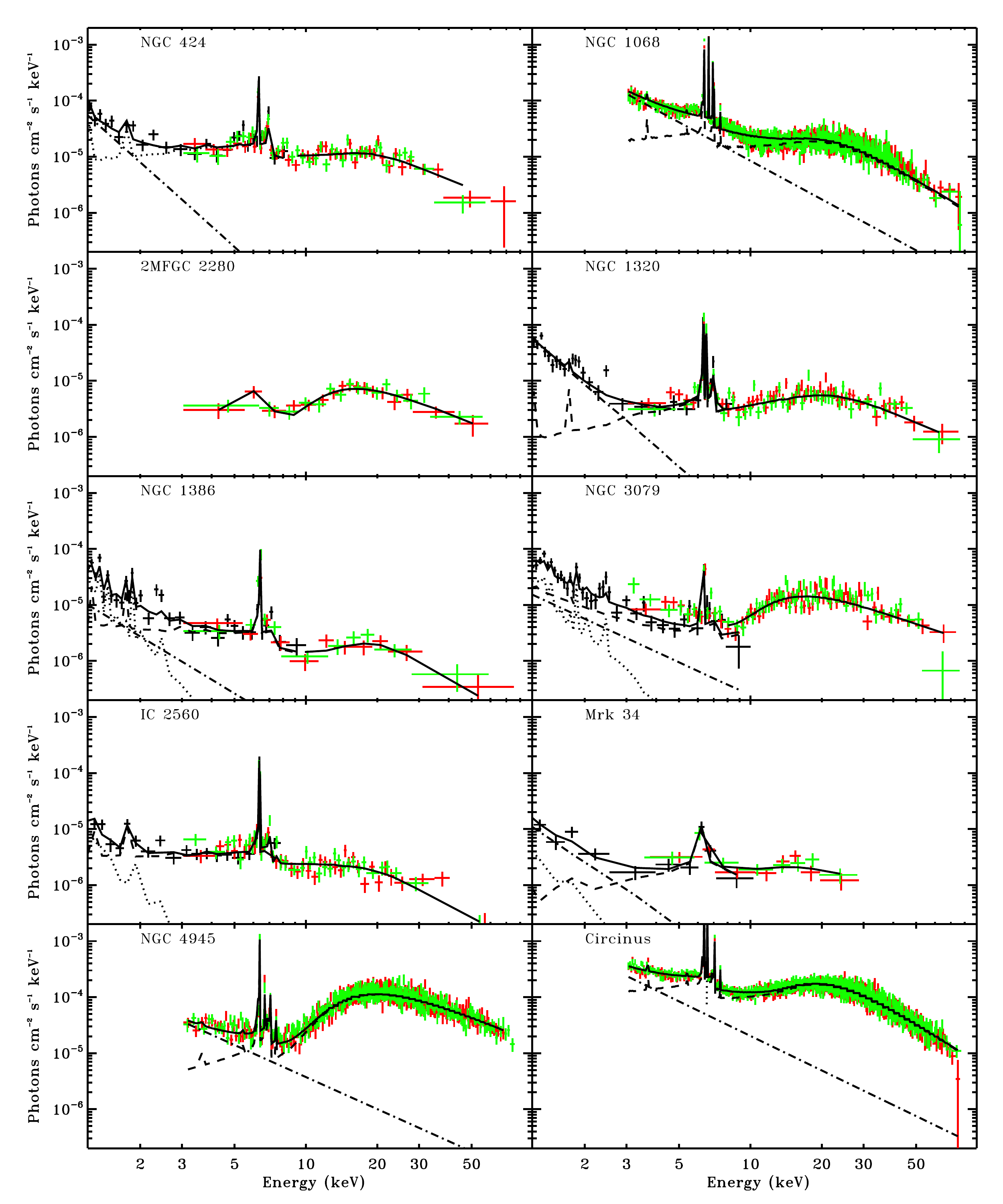}
\caption{Unfolded spectra for all sources fit with the {\sc torus} model, where \nustar\ FPMA data are shown in red, FPMB data are shown in green and \xmm\ data are shown in black. The solid black lines represent the sum of all model components, while the dashed line shows the {\sc torus} model. Dot-dashed lines represent the secondary power-law model and the dotted lines represent the thermal plasma model {\sc apec}, both used to fit the soft-X-rays. The spectra have been binned for plotting purposes, so that each bin has a detection of at least 3-$\sigma$ up to a maximum of 10 data points per bin.}
\label{fig_spec}
\end{center}
\end{figure*}

\begin{figure}
\begin{center}
\includegraphics[width=90mm]{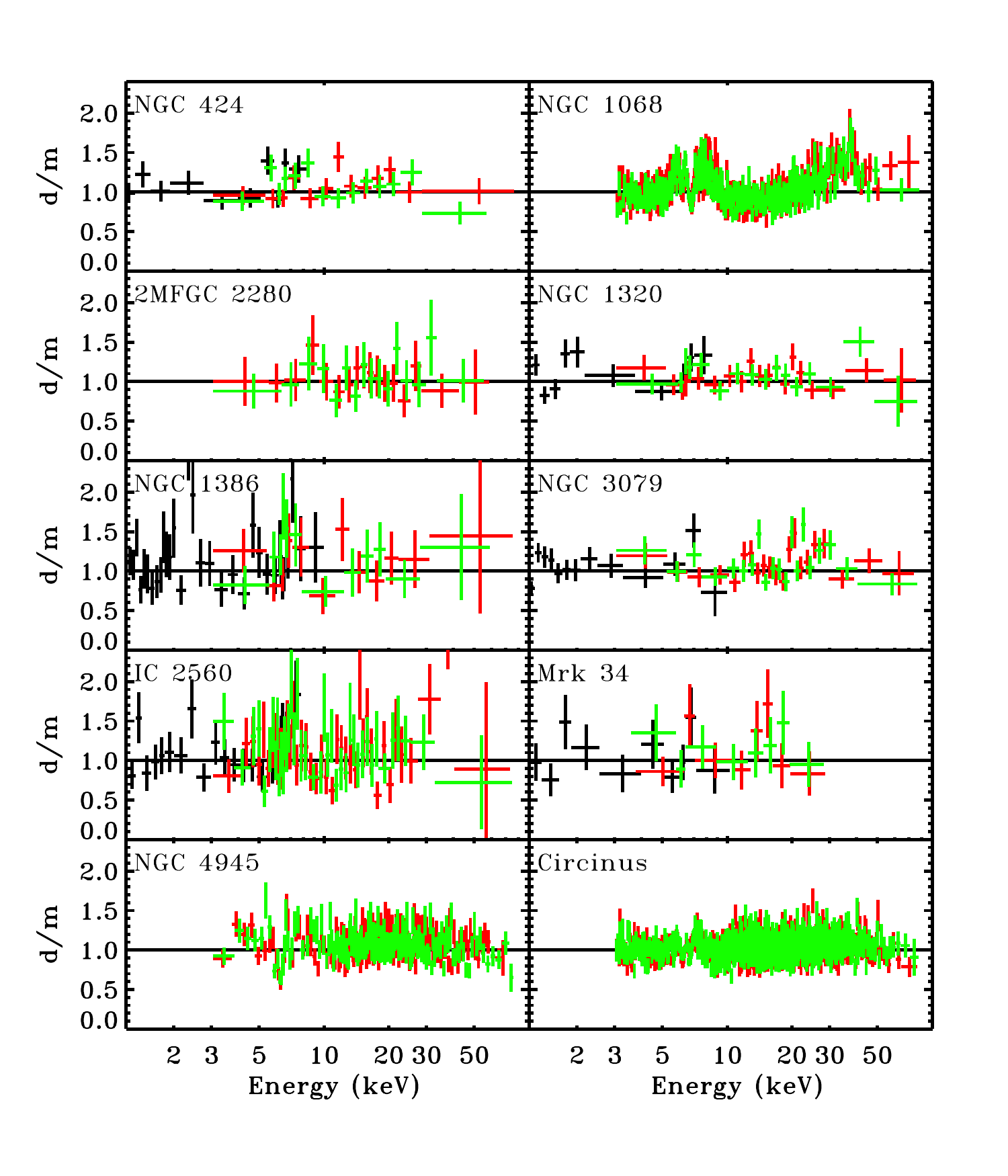}
\caption{Data-to-model ratios of the fits with the {\sc torus} model. The colors are the same as Fig. 5. The binning for this figure has been increased in order to better see deviations and trends in the residuals.} 
\label{fig_ratios}
\end{center}
\end{figure}

The best-fit unfolded spectra fit with the {\sc torus} model are presented in Figure \ref{fig_spec}. This illustrates the variety of spectral shapes seen above 10 keV for these Compton-thick sources, as well as the range in signal-to-noise in the sample. The data-to-model ratios are presented in Figure \ref{fig_ratios} which indicate how well the {\sc torus} model fits the data. The worst reduced \chisq\ from the {\sc torus} fits result from the spectrum of NGC~1068 (1.72), where there is significant curvature in the data-to-model ratio, implying that the {\sc torus} model is not a good description of the data. This was also the conclusion from the detailed analysis of this source in  \cite{bauer14}, which utilizes data from a large array of telescopes and several model combinations. They find that the data require a complex combination of models to reduce this curvature and that a monolithic torus structure is not likely.

\begin{figure*}
\begin{center}
\includegraphics[width=180mm]{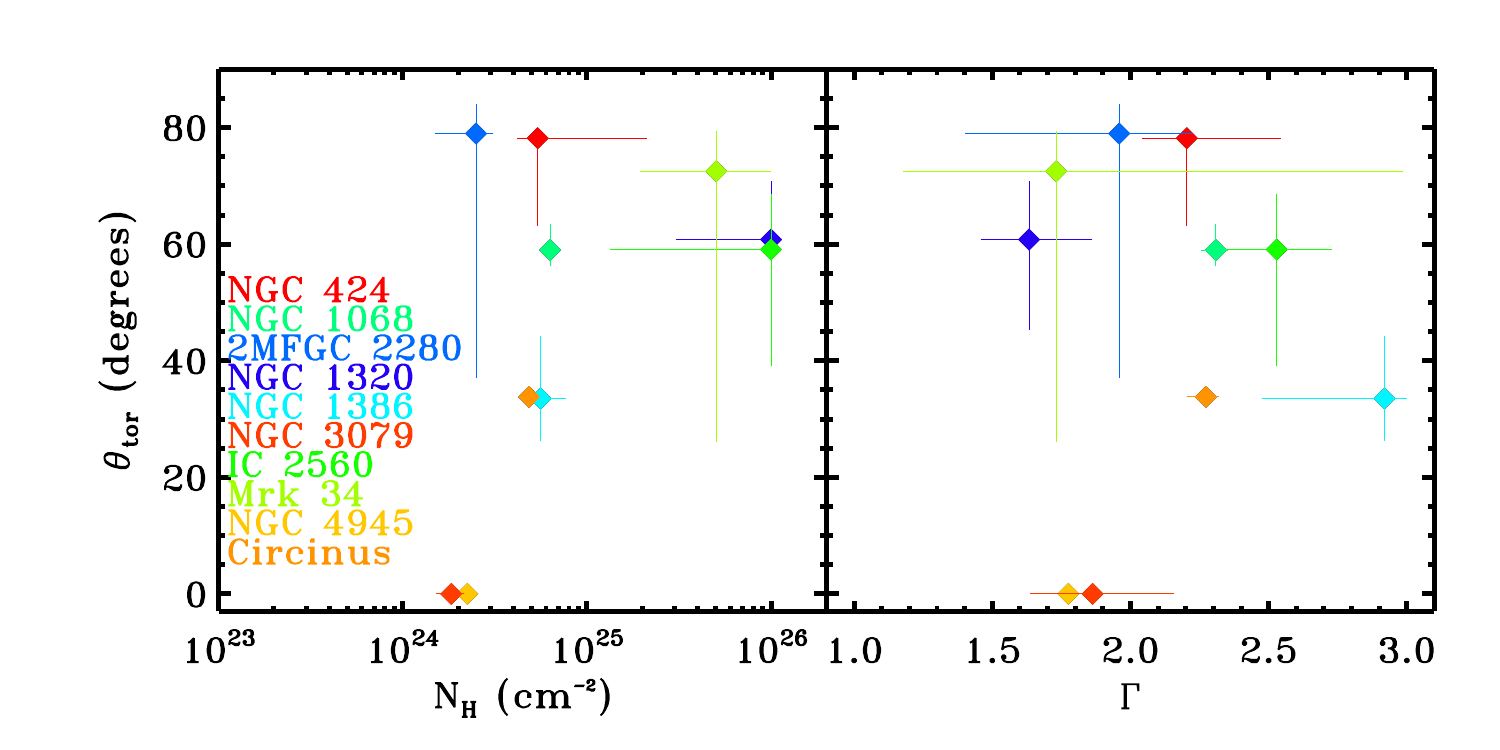}
\caption{Best-fit measurements of \thetator\ against best-fit measurements of \nh\ and $\Gamma$ from the {\sc torus} model. NGC~3079 and NGC~4945 are best-fit by the {\sc sphere} model implying that these sources have \thetator=0\degree. No uncertainties can be calculated for this, however, as the {\sc sphere} model has a fixed opening angle of zero.}
\label{fig_specpar}
\end{center}
\end{figure*}

First in our investigation into what drives the covering factor in AGN, we examine how the measurement of \thetator\ in our analysis relates to the other measurements made with the {\sc torus} model, namely \nh\ and $\Gamma$, by plotting these quantities against each other along with their uncertainties (Figure \ref{fig_specpar}). No uncertainties can be derived for the \thetator=0\degree\ values assigned to NGC~3079 and NGC~4945 as these have been determined from the {\sc sphere} model which has a fixed opening angle of zero. There is no clear relationship between these quantities.

\section{Potential biases and systematics}

In our analysis, we fix the torus inclination angle to an edge-on position of 87\degree. This is primarily motivated by the observation that the inclination angle does not have a large effect on the observed spectrum above 10 keV and thus allows us to reduce the number of free parameters in the fit. Furthermore an edge-on inclination angle allows the full range of opening angles to be explored. Fixing the inclination angle also avoids the scenario where the inclination angle approaches the opening angle, which produces a partially covered solution due to the angular binning in the {\sc torus} model. We investigate whether allowing the inclination angle to be free allows it to be constrained. This is not the case, however Circinus is an exception, as found by \cite{arevalo14}, thus we have allowed the inclination angle to be free for this source. 

The \nustar\ band is also well suited to studying the high-energy cut-off in the X-ray spectra of AGN, and has been used to constrain this in a number of bright unobscured sources \citep[e.g. SWIFT J2127.4+5654, IC 4329A, 3C 382, MCG-5-23-16,][Balokovi\'c et al. submitted, respectively]{marinucci14,brenneman14, ballantyne14}. However, the {\sc torus} model does not include this parameter and thus we have neglected its effect in this work. The potential effect of neglecting the high-energy cut-off here is to over-estimate the Compton hump, which is produced by the down-scattering of high-energy photons. As the covering factor roughly correlates with the strength of the Compton hump, then over-estimating the Compton hump will lead to a systematic under-estimation of the covering factor. We expect this effect to be very small as the lowest high-energy cut-off energy detected so far by \nustar\ is at 108 keV in SWIFT J2127.4+5654 \citep{marinucci14}, which is far above that of the Compton hump ($\sim30$ keV). For photons originating around a cut-off of this energy to affect the Compton hump, at least 10 scatterings would be required. To understand this effect fully, the {\sc torus} model should be improved to include the high-energy cut-off.

For this study, we have a heterogeneous sample in terms of X-ray spectral coverage and most of our sample have accompanying soft X-ray data to constrain the spectral components below 10 keV. However, for four sources, we use {\it NuSTAR} data alone. No soft X-ray data exist for 2MFGC~2280, meanwhile for NGC~1068, NGC~4945 and Circinus, the {\it NuSTAR} data are high signal-to-noise enough that they can constrain the {\sc torus} parameters alone. We test this assumption for all three sources by adding in an \xmm\ observation to model the soft X-rays below 3 keV. We find an additional {\sc apec} model is required to fit these data in addition to the {\sc torus} and power-law model already in place. Fitting the data simultaneously with the soft component constrained by \xmm\ produces only small changes in \thetator, and within measurement uncertainties. We conclude that the exclusion of the soft X-ray data does not bias our results significantly in these sources.

When fitting \xmm\ and \nustar\ data together, we have kept the spectral parameters fixed between the two data sets. Since the observations were not made simultaneously, spectral variability could be missed, with the exception of the intrinsic luminosities, which is accounted for by the cross-normalisation between the two observatories being a free parameter. We investigated spectral variability in the five sources where we use both \xmm\ and \nustar\ data together by allowing the \nh\ parameter to vary between data sets. We find that the \nh\ parameter is consistent within the uncertainties for all of these except NGC~424. For this source, \xmm\ measures \nh$=4.6^{+1.3}_{-0.7}\times10^{24}$ \cmsq\ and \nustar\ measures \nh$=2.1^{+0.6}_{-0.3}\times10^{24}$ \cmsq. Allowing for the \nh\ to vary like this does not however change our result on \thetator.

Lastly we investigate how further spectral components added to the fit may affect our results. Specifically, we investigate the addition of the semi-infinite slab reflection model {\sc pexrav} in order to represent a contribution to reflection from the accretion disc. We use the pure reflection component of this model and fix $\Gamma$ to that of the {\sc torus} model, set the high-energy cut-off to maximum ($1\times10^{6}$ eV), the adundances to solar and the cosine of the inclination angle to 0.45. The normalization is free. In four sources, NGC~1068, 2MFGC2280, NGC~1320, NGC~1386 and IC~2560, the normalization of the {\sc pexrav} component falls to zero in the fit. For NGC~3079 and NGC~4945, the addition of the {\sc pexrav} component has no effect on \thetator, whereas for Circinus, the change is $-5$\degree. For NGC~424 the addition of {\sc pexrav} to the fit causes a shift in \thetator\ of $-25$\degree. While the change in \thetator\ is large for NGC~424, when considering the  sample as a whole we conclude that the addition of this components does not appear to introduce a systematic effect in the determination of \thetator. We note that the {\sc pexrav} component added here is unabsorbed and not subjected to the same absorption as the primary power-law. To treat this correctly, the {\sc pexrav} spectrum would need to be added to the intrinsic emission of the Monte-Carlo models that calculate the effect of Compton scattering within the {\sc torus} model, which is not easily done.

\section{Comparison to previous results}

The use of the {\sc torus} model here to determine the opening angle of the torus is fairly novel, as is the use of X-ray torus models to understand high-energy emission in heavily obscured AGN. Therefore, in Section \ref{sec_models} we compared the spectral fit parameters determined from the classical {\sc pexrav} model and the more recent {\sc mytorus} model when fitted to simulated spectra from the {\sc torus} model for a range of \thetator\ and \nh. Here, we compare the fits to real \nustar\ data with the {\sc torus} model presented here to those made with {\sc pexrav} and {\sc mytorus} from previous published works. We list the different \nh\ and $\Gamma$ values obtained from each in Table \ref{table_speccomp}. 

In our initial analysis, we found that {\sc pexrav} systematically underestimates $\Gamma$ with respect to both {\sc torus} and {\sc mytorus}, most likely because it cannot reproduce the full strength of the Compton hump produced by a torus geometry. This is also found in our comparison of fits to real data, as fits with {\sc pexrav} produce a lower $\Gamma$ than both the {\sc torus} or {\sc mytorus} model for NGC~424, NGC~1068, NGC~1320 and IC~2560. 

As for fits with {\sc mytorus} with respect to the {\sc torus} model, the $\Gamma$ measurements are consistent with each other within the uncertainties, with the exceptions of NGC~1068 and NGC~4945. For NGC~1068,  \cite{bauer14} use a large, combined set of X-ray spectral data from different observatories with far more detailed spectral modelling than is done here, including modelling of the host spectrum. Futhermore, the {\sc mytorus} fits are done in decoupled mode. For NGC~4945, the {\sc torus} model measures $\Gamma$ $\sim$0.2 lower than {\sc mytorus}. However, the measurement of $\Gamma$ using {\sc mytorus} in \cite{puccetti14} was done on a time-resolved basis with {\sc mytorus} in decoupled mode, where the parameters of the reflection components are not fixed to those of the transmitted component. However, we do measure a lower \nh\ for NGC~4945 than the {\sc mytorus} model, which may be the reason for the discrepancy. Furthermore, a fit with the {\sc sphere} model produces a better fit to the data and a $\Gamma$ value consistent with {\sc mytorus} (1.78$^{+0.13}_{-0.06}$). For Mrk 34, the difference in the measured $\Gamma$ values between the two models is $\sim0.5$, with {\sc mytorus} producing the steeper slope. The {\sc torus} model, however, measures \nh$>10^{25}$ \cmsq\ in Mrk 34, whereas {\sc mytorus} is limited to \nh$<10^{25}$ \cmsq, and, indeed the upper limit is reached in the fit. The inability of {\sc mytorus} to measure columns greater than this is the likely cause of disagreement in Mrk 34, as also noted by \cite{gandhi14}. The two measured values are nevertheless consistent within the large uncertainties in $\Gamma$.

\begin{table*}
\centering
\caption{Comparison of fits with the {\sc torus} model to previous analyses with {\sc pexrav} and {\sc mytorus}}
\label{table_speccomp}
\begin{center}
\begin{tabular}{l l l l l l l l}
\hline
Source name & \nh\ ({\sc mytorus}) & \nh\ ({\sc torus}) & $\Gamma$ ({\sc pexrav}) & $\Gamma$ ({\sc mytorus}) & $\Gamma$ ({\sc torus}) & $\theta_{\rm tor}$\\
(1) & (2) & (3) & (4) & (5) & (6) & (7) \\
\hline

NGC 424 & 3$\pm$1 &   5.41$^{+  15.67}_{-   1.23}$ &  1.66$\pm$0.09 & 2.07$^{+ 0.11}_{- 0.09}$ & 2.20$^{+ 0.34}_{- 0.16}$ & 78.19$^{+ 0.43}_{-15.01}$ \\
NGC 1068 & 10$^{+u}_{-4.4}$ &   6.32$^{+   0.44}_{-   0.05}$ & 1.57$\pm$0.02 & 2.20$^{+0.02}_{-0.01}$ & 2.31$^{+ 0.04}_{- 0.05}$ & 58.99$^{+ 4.51}_{- 2.76}$ \\
NGC 1320 &  4$^{+4}_{-2}$ & 100.00$^{+u}_{-  70.81}$ &  1.3$\pm$0.1 & 1.6$\pm$0.2 & 1.66$^{+ 0.21}_{- 0.19}$ & 60.17$^{+10.69}_{-14.91}$ \\
IC 2560 & 10$^{+u}_{-3}$ & 100.00$^{+u}_{-  86.64}$ & 2.2$^{+ 0.1}_{- 0.2}$ & 2.55 (f) & 2.53$^{+ 0.20}_{- 0.20}$ & 59.08$^{+ 9.57}_{-20.05}$ \\
Mrk 34 & 2.45$^{+u}_{-1.08}$ &  50.43$^{+  49.57}_{-  31.01}$ & - & 2.2$^{ +0.2}_{- 0.3}$ & 1.73$^{+ 1.25}_{- 0.56}$ & 72.51$^{+ 6.89}_{-46.51}$ \\
NGC 4945 &  3.5$\pm$0.1 &   2.54$^{+   0.13}_{-   0.15}$ & - & 1.77-1.96 & 1.58$^{+ 0.04}_{- 0.04}$ & 26.00$^{+ 0.50}_{-l}$ \\
Circinus & 10.0$\pm$1.8 &   4.85$^{+   0.39}_{-   0.42}$ & - & 2.34$\pm$0.02 & 2.27$^{+ 0.05}_{- 0.07}$ & 33.79$^{+ 1.83}_{- 1.56}$ \\

\hline
\end{tabular}
\tablecomments{From {\protect\cite{balokovic14}} for NGC~424, NGC~1320 and IC~2560,  {\protect\cite{bauer14}} for NGC~1068, {\protect\cite{gandhi14}} for Mrk 34, {\protect\cite{puccetti14}} for NGC~4945 and {\protect\cite{arevalo14}} for Circinus. Column (1) lists the source name, column (2) gives the \nh\ measured by {\protect\sc mytorus} in units of $10^{24}$ \cmsq. +u indicates that the upper constraint on the \nh\ from {\sc mytorus} is beyond the upper limit of the model. Column (3) gives the \nh\ measured by the {\sc torus} model in the same units, column (4) gives the photon index measured by {\sc pexrav}, column (5) gives the photon-index measured by {\sc mytorus}, where (f) indicates this parameter has been fixed, and column (6) gives the photon index measured by {\sc torus}. Column (7) lists \thetator\ measured by the {\sc torus} model in units of degrees. $+u$ indicates that a parameter has reached the upper limit when estimating the uncertainty. $-l$ indicates that the lower limit has been reached.}
\end{center}
\end{table*}

\subsection{NGC~4945}

The high covering factor determined here for NGC~4945 is in disagreement with previous analyses which conclude that NGC~4945 has a small covering factor, of order 0.1 \citep{madejski00, done03, yaqoob12, puccetti14}. This conclusion was drawn from the fact that NGC~4945 is variable above 10 keV and that it has a relatively weak reflected component. \cite{madejski00} used Monte-Carlo simulations to show that the fraction of unscattered photons reaching the observer is much higher for low covering factors (63\% for \thetator=80\degree) than high covering factors (19\% for \thetator=10\degree). The argument states that with an optical depth to Compton scattering of 2-3, the intrinsic variability of the AGN with a high covering factor would be smeared out by reflection and that a low covering factor would be required to produce such weak reflection. 

\cite{done03} also conclude that the Compton-thick material cannot cover the whole source in NGC~4945 as the 6.4 keV emission is spatially extended in {\it Chandra} observations and must be illuminated by hard X-rays from the AGN. \cite{done03} also suggest that NGC~4945 must be fully covered due to the lack of high ionization optical/IR lines, although this material does not need to be Compton-thick. 

\cite{yaqoob12} presented an extensive broad-band X-ray spectral analysis of NGC~4945 using {\it Suzaku}, {\it BeppoSAX} and {\it Swift}/BAT data using {\sc mytorus}, {\sc torus} and {\sc sphere} models and also prefer the small covering factor solution due to the hard X-ray variability.

A detailed spectral and temporal analysis of NGC~4945 with \nustar\ was presented in \cite{puccetti14} utilizing the {\sc mytorus} model in decoupled mode. The hard X-ray variability was confirmed, where variations of a factor of two above 10 keV were reported. They estimate the covering factor for this source, using the ratio of the reflected component to the direct component. Since this is very small, they conclude that the covering factor is $\sim$0.13, and fairly constant with flux. 

Using the {\sc torus} model, we have found that the X-ray spectral shape of NGC~4945 can also be produced by a source with a high covering factor of obscuring Compton thick material and that for this model this in fact provides a better fit to the X-ray spectrum than the low covering factor scenario. Figure \ref{fig_thetacon} shows \chisq\ as a function of \thetator\ for the fit to NGC~4945. Although this shows a local minimum at large opening angles (low covering factor), the minimum at small opening angles (large covering factor) is significantly lower.  Indeed, with their Monte-Carlo simulations, \cite{madejski00} using  {\it RXTE} data also find that the spectral shape favours a small opening angle, finding \chisq=68.5 for \thetator=20\degree\ and \chisq=75.4 for \thetator=80\degree. Their conclusion regarding the large opening angle is instead driven by the variability and their Monte-Carlo simulations. However, it is not clear whether these results based on the variability are dependent on the geometry of the torus used (a torus with a square cross-section), or if these results are energy dependent. 

Our results also imply an optical depth to scattering of 1.5--1.7, which is lower than previously considered, and thus the effect of scattering is slightly diminished. However the fraction of directly transmitted photons is still low when considering a high covering factor, which is hard to reconcile with the hard X-ray variability. 

The conclusions regarding the covering factor of Compton-thick material surrounding NGC~4945 drawn from the variability and those drawn from the spectral shape are at odds and the models used to draw these conclusions both have their limitations. Concerning the variability, it has been assumed that scattered photons cannot transmit the intrinsic variability of the source. However, Compton scattering prefers forward (and backward) scattering with small angles, so the difference in light travel time between transmitted and scattered photons need not be large. The difference in light travel time is also dependent on the distance of the scatterer from the central source. For the {\sc torus} model, the analysis is limited to a range of covering factors, not allowing investigations of very small covering factors ($\sim0.1$) or very high covering factors ($\sim0.9-0.99$), and also does not allow one to decouple the transmitted and scattered components. The assumption of a smooth matter distribution is also unlikely to be accurate as it is most likely to be clumpy \citep{yaqoob12}. It is clear that further work is required to fully understand the nature of the absorber in NGC~4945.

\section{What determines the covering factor of the obscurer in AGN?}
\label{sec_determines}

The obscurer in local AGN is widely regarded to be a cold molecular torus, which many recent results imply has a clumpy constituency \citep[e.g.][]{elitzur06, markowitz14}. As discussed, high-energy X-rays are ideally suited to the study of the obscuring material, since Compton scattering effects from the gas in the obscuring medium dominate in this regime ($>$10 keV). Our fits to \nustar\ data with the {\sc torus} model, which assumes a smooth torus, support the general torus paradigm. In our analysis we determine a wide range of \thetator\ allowed by the model, and furthermore identify three sources where the spectral fits indicate small or zero \thetator. 

\begin{figure*}
\begin{center}
\includegraphics[width=160mm]{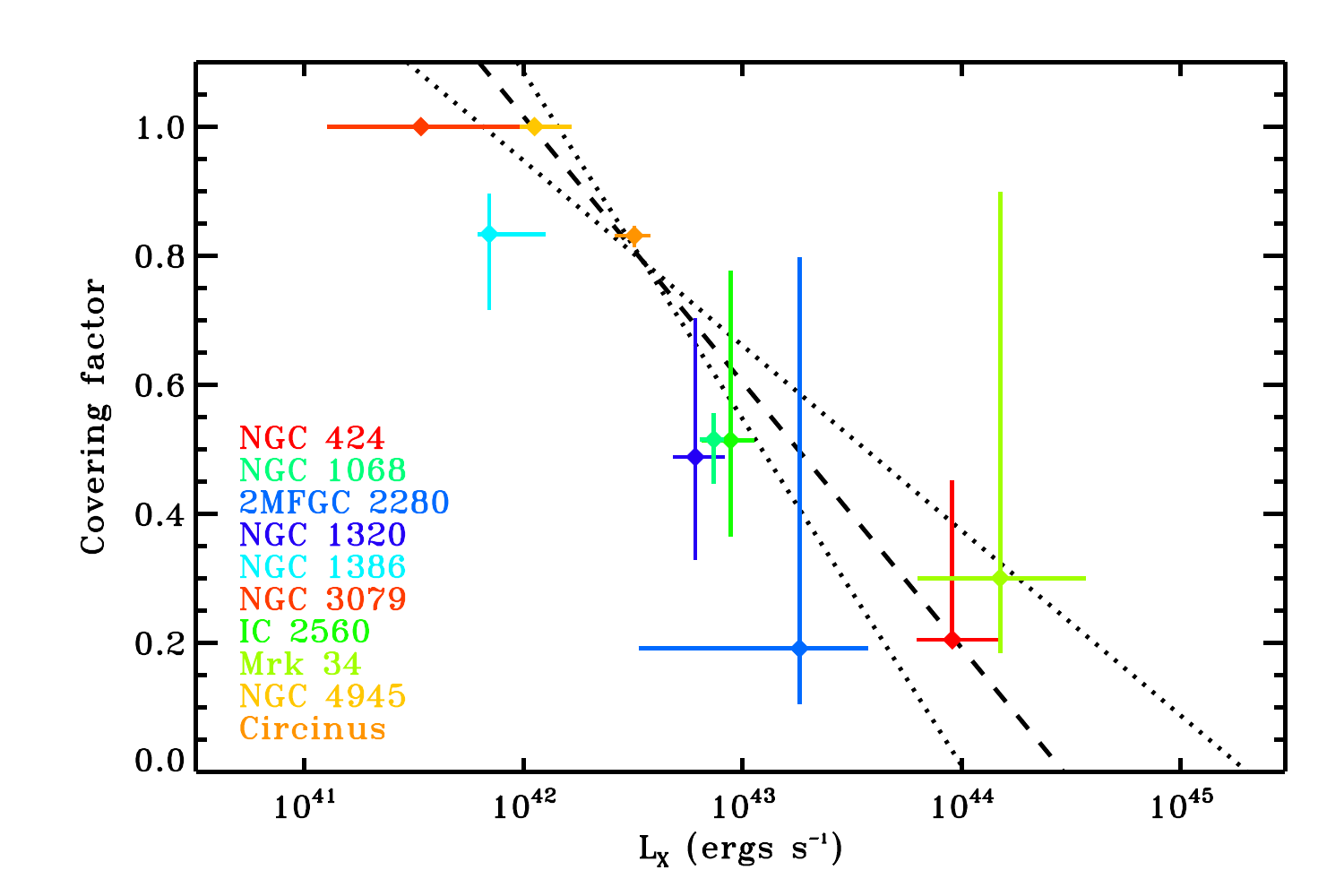}
\caption{The covering factor as a function of intrinsic 2-10 keV luminosity, \lx, derived in our analysis, where the best-fit linear model $f_{\rm c}=(-0.41\pm0.13)$log$_{10}$(\lx/\ergs)$+18.31\pm5.33$ is plotted. NGC~3079 and NGC~4945 are excluded from the linear function fit since they both have no uncertainties on their covering factor. Nonetheless, they both agree well with the derived function.}
\label{fig_fobs1}
\end{center}
\end{figure*}

Although our sample is small, we investigate what could be physically influencing the opening angle. As the obscured fraction is known to depend on X-ray luminosity, we first investigate how the covering factors derived here depend on \lx, using the 2-10 keV band and the intrinsic luminosity determined from the model. Figure \ref{fig_fobs1} plots these quantities against each other with their measured uncertainties. A strong anti-correlation is seen, as expected, where one of the most luminous sources, NGC~424 with log$_{10}$(\lx/\ergs)=43.96$^{+ 0.21}_{- 0.16}$ has a small, relatively well constrained covering factor (\fcov=0.20$^{+ 0.25}_{- 0.00}$), while Circinus with a moderate luminosity of log$_{10}$(\lx/\ergs)=42.51$^{+ 0.07}_{- 0.09}$ has a larger covering factor of \fcov=0.83$^{+ 0.01}_{- 0.02}$. Such a correlation is expected due to more luminous sources sublimating dust in the inner edge of the torus at larger distances. For the same vertical extension of the torus, a larger radius of the inner part of the torus gives a lower covering factor of the central source due to geometrical effects.

We fit a simple linear model, $y=mx+c$ to the covering factor vs. log$_{10}$\lx\ data using the IDL function {\sc linfit}, which utilises \chisq\ minimisation and takes into account the uncertainties in the covering factor. We find that $f_{\rm c}=(-0.41\pm0.13)$log$_{10}$(\lx/\ergs)$+18.31\pm5.33$, where the uncertainties are 1-$\sigma$. We plot  this function along with the uncertainties in Figure \ref{fig_fobs1}. We do not include the data for NGC~3079 or NGC~4945 in the fit as these have no uncertainties in the covering factors assigned to them. Notably, however, they both agree very well with the fitted model.

\begin{figure*}
\begin{center}
\includegraphics[width=160mm]{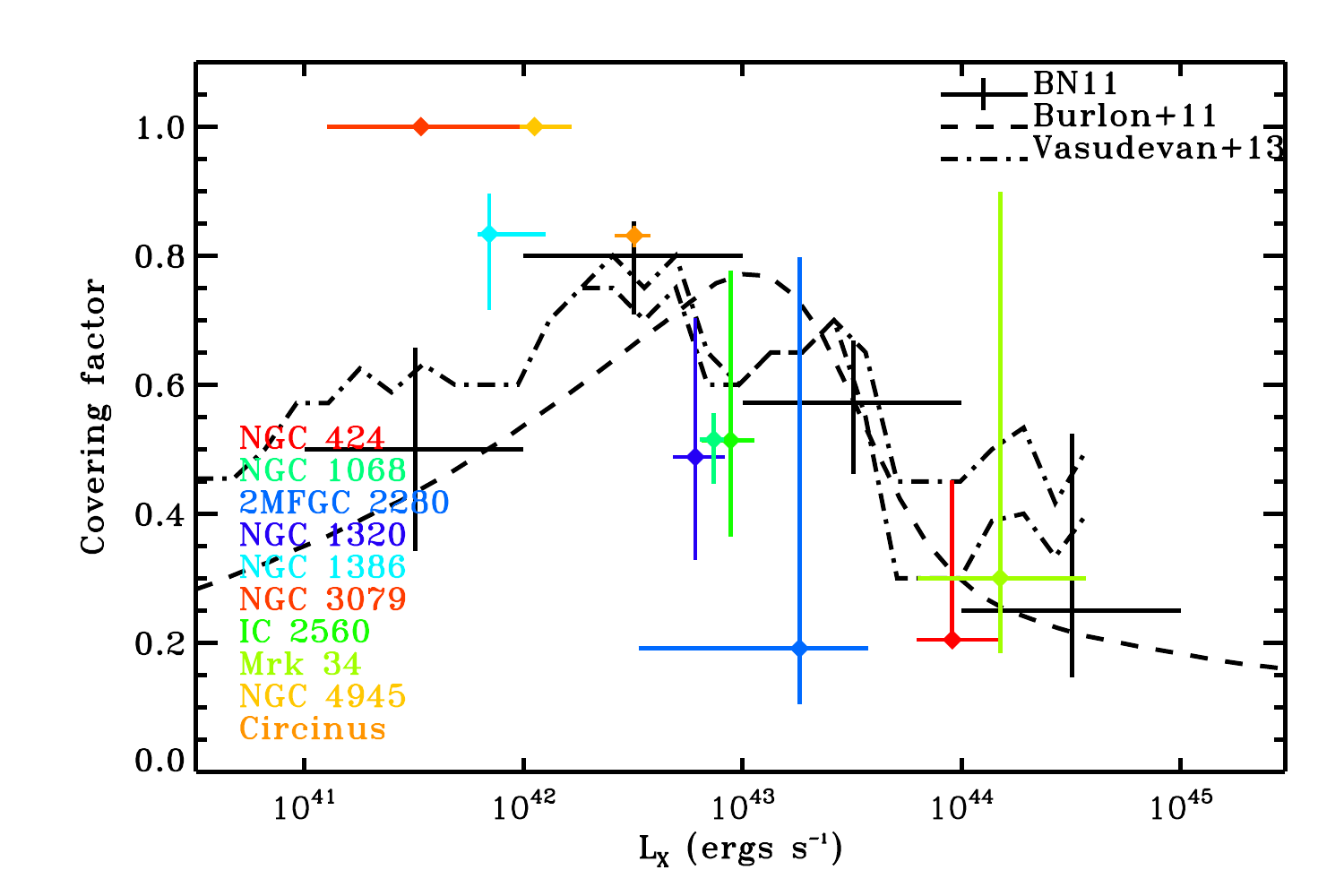}
\caption{The covering factor of the torus as a function of intrinsic 2-10 keV luminosity, \lx, derived in our analysis in comparison to the obscured fraction of local AGN determined from Burlon et al. (2011), Brightman \& Nandra (2011) and Vasudevan et al. (2013). The covering factors agree well with the obscured fraction for \lx$\gtrsim10^{42.5}$ \ergs. However, at low luminosities, we find no evidence for the decline in the covering factor seen in those studies.}
\label{fig_fobs}
\end{center}
\end{figure*}

We also compare the covering factors to the obscured fraction of local AGN from three studies. We compare to recent obscured fractions presented by \cite{burlon11} and \cite{vasudevan13}, both hard X-ray selected samples from {\it Swift}/BAT, and that of \cite{brightman11b}, a mid-infrared, {\it IRAS}-selected sample. All three obscured fractions are defined as the fraction of sources with \nh$>10^{22}$ \cmsq, calculated in different luminosity bins. We plot these obscured fractions in Figure \ref{fig_fobs}. The uncertainties in the \cite{brightman11b} data points are binomial. The \cite{burlon11} curve is calculated by dividing the X-ray luminosity function (XLF) of obscured AGN by the total XLF, done in the 15-55 keV band. The \cite{vasudevan13} line is a running average using 30 sources per bin. The two different lines for this sample are derived from where there are uncertainties on the \nh\ measurement, and the upper \nh\ bound is used for the upper line and the lower \nh\ bound is used for the lower line.

 The obscured fraction determined in these three studies agree very well with each other despite the differing selections and determinations, declining from a peak at \lx$=10^{42-43}$ \ergs\ towards higher luminosities. All three studies also find evidence for a decline in the obscured fraction towards lower luminosities.

Also in Figure \ref{fig_fobs}, we overplot the covering factors derived from our sample. We find good agreement between our derived covering factors and the obscured fraction for \lx$\gtrsim10^{42.5}$ \ergs. However, while previous studies have found evidence for a decrease in the covering factor at low luminosities, we find that our sources in the \lx$=10^{41-42}$ \ergs\ range are heavily buried in material with high covering factors. A larger more complete sample is required to show if this disagreement is statistically significant or due to the low number statistics of our sample.

We note that the covering factors derived here are those of the Compton-thick gas surrounding the AGN, since the {\sc torus} model assumes a constant density torus with a constant \nh\ as a function of inclination angle. Therefore any additional covering by Compton-thin gas would not be recognised, in which case the covering factors here could underestimate the total covering factor. Due to the agreement between the Compton-thick covering factor and the obscured fraction (for \lx$>10^{42.5}$ \ergs), this does not appear to be the case in our sample. This implies uniform torus covering factors for Compton-thin and Compton-thick AGN given the same \lx, above 10$^{42.5}$ \ergs. Best estimates of the local Compton-thick fraction put it at $\sim20$\% \citep{burlon11,brightman11b}, which is lower than the covering factors determined here, which could suggest that a larger population of CT AGN exists in the local Universe.

\begin{figure*}
\begin{center}
\includegraphics[width=180mm]{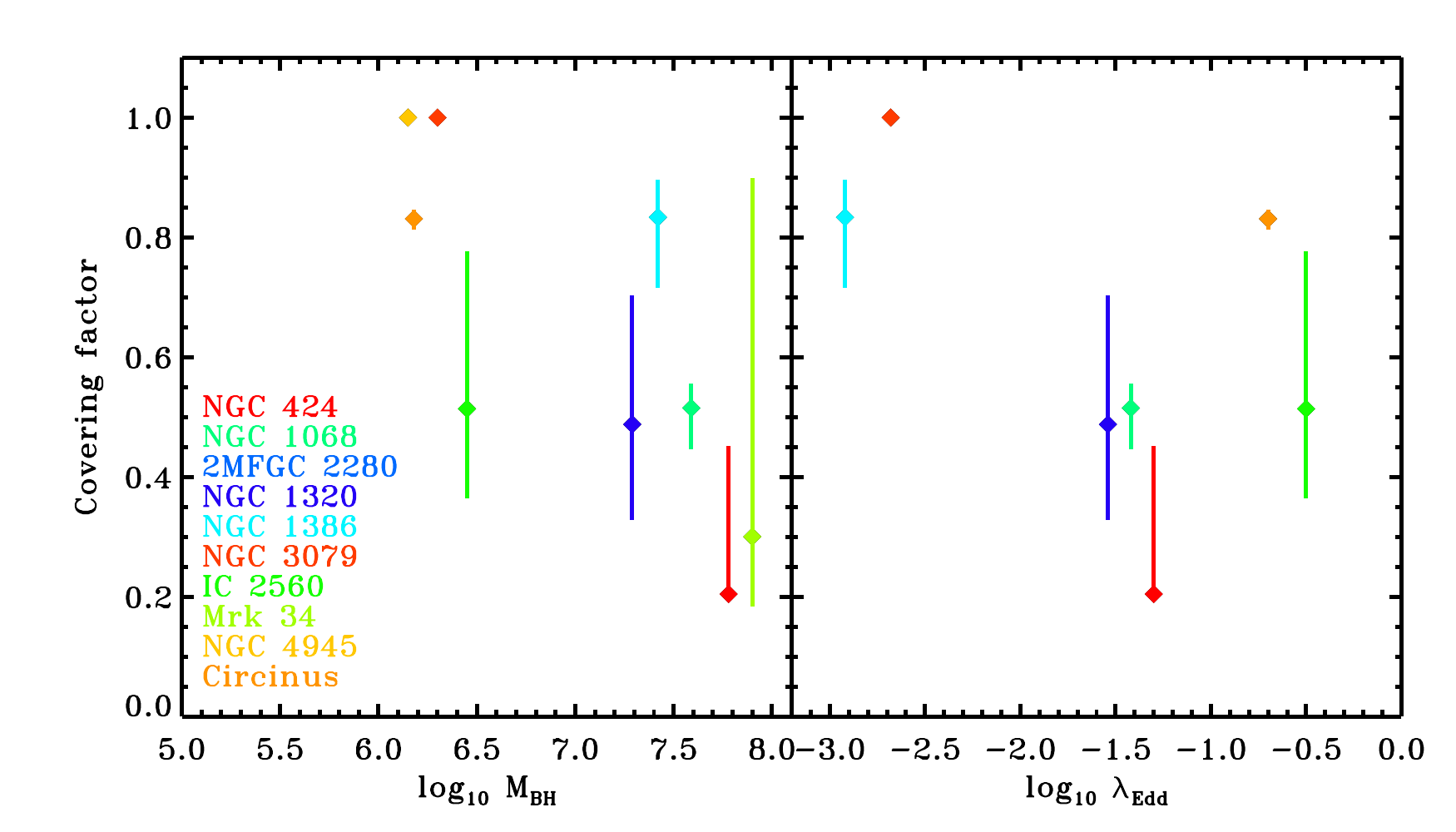}
\caption{Our derived covering factor of the obscuring material compared to the black hole masses (in solar masses) and the Eddington ratios from the literature.}
\label{fig_accrpar}
\end{center}
\end{figure*}

Lastly, we briefly investigate what other AGN parameters may be involved in determining the covering factor of the obscurer. We have already shown that the X-ray luminosity, which traces the bolometric power of the AGN and is thus dependent on the mass accretion rate, plays an important role. We next explore if the mass of the black hole, \mbh, or the fraction of the Eddington luminosity, \lamedd, physically influences the covering factor. These quantities are notoriously difficult to determine in obscured AGN, as virial mass estimates from optical broad lines are not accessible. In these cases, the velocity dispersion of the stars in the bulge is often used to estimate \mbh\ from the \mbh$-\sigma_{*}$ relation, although there is evidence for large scatter in this relationship, especially at low mass \citep{greene10}. For some sources, water megamasers can provide robust black hole mass measurements. We assemble this data from the literature and list it in Table \ref{table_accrpar}. Some \lamedd\ estimates are also available. The covering factor of the torus relative to these quantities are plotted in Figure \ref{fig_accrpar}. 

This preliminary investigation seems to show that the highest black hole mass systems in our sample have the smallest covering factor, while the smallest black holes have the highest covering factors, however the relationship is not statistically significant and a far larger sample is required to confirm this trend and to break the degeneracy with \lx. As for the covering factor as a function of \lamedd, our data suggest that high covering factors are exhibited in both low ($10^{-3}$) and high ($\sim0.3$) \lamedd\ systems, in line with the conclusions of \cite{draper10} who find that CTAGN are made up of a composite population of both high ($>$0.9) and low ($<$0.01) Eddington ratio systems.

\begin{table*}
\centering
\caption{Accretion parameters of our sample}
\label{table_accrpar}
\begin{center}
\begin{tabular}{l c c c c c c}
\hline
Source name & class & \fcov\ & log$_{10}$\lx\ & log$_{10}$\mbh\ & log$_{10}$\lamedd\ & ref  \\
(1) & (2) & (3) & (4) & (5) & (6) & (7)  \\
\hline

NGC 424& Sy1h &  0.20$^{+ 0.25}_{- 0.01}$ & 43.96$^{+ 0.21}_{- 0.16}$ &  7.78 & -1.30 &  1, 6 \\
NGC 1068& Sy1h &  0.52$^{+ 0.04}_{- 0.07}$ & 42.87$^{+ 0.04}_{- 0.06}$ &  7.59 & -1.42 &  2, 2 \\
2MFGC 2280& Sy2 &  0.19$^{+ 0.61}_{- 0.09}$ & 43.26$^{+ 0.31}_{- 0.74}$ &  0.00 &-99.00 &  0, 0 \\
NGC 1320& Sy2 &  0.50$^{+ 0.21}_{- 0.17}$ & 42.79$^{+ 0.12}_{- 0.09}$ &  7.29 & -1.54 &  2, 2 \\
NGC 1386& Sy1i &  0.83$^{+ 0.06}_{- 0.12}$ & 41.84$^{+ 0.26}_{- 0.05}$ &  7.42 & -2.92 &  2, 2 \\
NGC 3079& Sy2 &  1.00$^{+ 0.16}_{- 0.03}$ & 41.96$^{+ 0.19}_{- 0.33}$ &  6.30 & -2.68 &  7, 2 \\
IC 2560& Sy2 &  0.51$^{+ 0.26}_{- 0.15}$ & 42.95$^{+ 0.11}_{- 0.13}$ &  6.45 & -0.50 &  3, 6 \\
Mrk 34& Sy2 &  0.30$^{+ 0.60}_{- 0.12}$ & 44.18$^{+ 0.39}_{- 0.38}$ &  7.90 &-99.00 &  5, 0 \\
NGC 4945& FSRS &  1.00$^{+ 0.00}_{- 0.00}$ & 41.92$^{+ 0.07}_{- 0.08}$ &  6.15 &-99.00 &  4, 0 \\
Circinus& Sy1h &  0.83$^{+ 0.01}_{- 0.02}$ & 42.51$^{+ 0.07}_{- 0.09}$ &  6.18 & -0.70 &  8, 9 \\

\hline
\end{tabular}
\tablecomments{Column (1) lists the source name, column (2) gives the AGN activity class based on the optical spectrum, where Sy1h indicates a Seyfert 2 with a hidden broad line region revealed in polarised light, Sy1i indicates a Seyfert 2 with a broad line region visible in the infrared, Sy2 indicates a Seyfert 2 with no evidence for a hidden broad line region and FSRS means a flat spectrum radio source. Column (3) gives the covering factor of the obscuring material derived in this work, column (4) gives log$_{10}$(\lx/\ergs) also derived here, column (5) lists the black hole masses from the literature in logarithm of solar masses, column (6) gives the Eddington ratio published in the literature and column (7) list the references for these data: 1. {\protect\cite{bian07}} (black hole mass from stellar velocity dispersion),  2. {\protect\cite{marinucci12}} (black hole mass from stellar velocity dispersion) 3. {\protect\cite{ishihara01}} (black hole mass from water megamaser emission), 4. {\protect\cite{greenhill97}} (black hole mass from water megamaser emission), 5. {\protect\cite{su08}} (black hole mass from [OIII] line width), 6. {\protect\cite{balokovic14}}; 7. {\protect\cite{kondratko05}}; 8. {\protect\cite{greenhill03}} (black hole mass from water megamaser emission); 9. {\protect\cite{arevalo14}}. }
\end{center}
\end{table*}

\section{Conclusions}

In this paper we have used the X-ray torus models of \cite{brightman11} and data from \nustar\ and \xmm\ to determine the covering factor of the Compton-thick gas in ten local CT AGN, NGC~424, NGC~1068, 2MFGC~2280, NGC~1320, NGC~1386, IC~2560, Mrk 34, NGC~3079, NGC~4945 and Circinus. We have also assessed the differences between the {\sc torus} model, {\sc pexrav} and {\sc mytorus}. We find:

\begin{itemize}

\item The slab reflection model, {\sc pexrav}, does not easily reproduce the Compton hump shape produced by a torus geometry, under-producing it for large covering factors, and over-producing it for small ones, resulting in a systematic offset in the parameters obtained. We therefore discourage use of this model in the fitting of high-energy X-ray emission of CTAGN. Our results compare well with {\sc mytorus} for \nh$<10^{25}$ \cmsq, where that model is valid, however we support the use of the covering factor as a free parameter in torus models.

\item Measurements of \thetator\ are in the range of 26--80\degree, limited by the allowed range of the model, with uncertainties on these measurements ranging from 5--40\degree. These correspond to covering factors in the range of 0.2--0.9, with uncertainties ranging from 0.05-0.6.

\item The covering factor is a strongly decreasing function of intrinsic 2-10 keV luminosity; when fitted with a linear function, we find $f_{\rm c}=(-0.41\pm0.13)$log$_{10}$(\lx/\ergs)$+18.31\pm5.33$.

\item The individual covering factors derived here agree well with the average covering factor of local AGN as measured by the obscured fraction as a function of \lx\ above 10$^{42.5}$ \ergs. However, while previous studies have found evidence for a decrease in the covering factor at low luminosities, we find that our sources in the \lx$=10^{41-42}$ \ergs\ range are heavily buried in material with high covering factors. A larger more complete sample is required to show if this disagreement is statistically significant or due to the low number statistics of our sample.

\item We find a conflicting result on NGC~4945, where our spectral analysis implies a large, almost 100\%, covering factor, whereas previous results have concluded that this source has a very low covering factor due to flux variability above 10 keV. We conclude that model limitations in both cases are the likely cause of the disagreement.

\end{itemize}

\acknowledgments

This work was supported under NASA Contract No. NNG08FD60C, and made use of data from the {\it NuSTAR} mission, a project led by the California Institute of Technology, managed by the Jet Propulsion Laboratory, and funded by the National Aeronautics and Space Administration. We thank the {\it NuSTAR} Operations, Software and Calibration teams for support with the execution and analysis of these observations.  This research has made use of the {\it NuSTAR} Data Analysis Software (NuSTARDAS) jointly developed by the ASI Science Data Center (ASDC, Italy) and the California Institute of Technology (USA). The work presented here was also based on observations obtained with XMM-Newton, an ESA science mission with instruments and contributions directly funded by ESA Member States and NASA. This research has also made use of data and software provided by the High Energy Astrophysics Science Archive Research Center (HEASARC), which is a service of the Astrophysics Science Division at NASA/GSFC and the High Energy Astrophysics Division of the Smithsonian Astrophysical Observatory. Furthermore, this research has made use of the NASA/IPAC Extragalactic Database (NED) which is operated by the Jet Propulsion Laboratory, California Institute of Technology, under contract with the National Aeronautics and Space Administration. DRB acknowledges support from NSF award AST 1008067; P.G. acknowledges support from STFC (grant reference ST/J003697/1), M. K. acknowledges support from the Swiss National Science Foundation (SNSF) through the Ambizione fellowship grant PZ00P2\textunderscore154799/1. We also acknowledge support from CONICYT-Chile grants Basal-CATA PFB-06/2007 (FEB), FONDECYT 1141218 (FEB), "EMBIGGEN" Anillo ACT1101 (FEB), Project IC120009 "Millennium Institute of Astrophysics (MAS)" of the Iniciativa Cient\'{\i}fica Milenio del Ministerio de Econom\'{\i}a, Fomento y Turismo (FEB).

{\it Facilities:} \facility{\nustar}, \facility{\xmm\ (EPIC)}


\begin{thebibliography}{}
\expandafter\ifx\csname natexlab\endcsname\relax\def\natexlab#1{#1}\fi

\bibitem[{{Akylas} \& {Georgantopoulos}(2009)}]{akylas09}
{Akylas}, A., \& {Georgantopoulos}, I. 2009, \aap, 500, 999

\bibitem[{{Akylas} {et~al.}(2006){Akylas}, {Georgantopoulos}, {Georgakakis},
  {Kitsionas}, \& {Hatziminaoglou}}]{akylas06}
{Akylas}, A., {Georgantopoulos}, I., {Georgakakis}, A., {Kitsionas}, S., \&
  {Hatziminaoglou}, E. 2006, \aap, 459, 693

\bibitem[{{Anders} \& {Grevesse}(1989)}]{anders89}
{Anders}, E., \& {Grevesse}, N. 1989, \gca, 53, 197

\bibitem[{{Ar{\'e}valo} {et~al.}(2014){Ar{\'e}valo}, {Bauer}, {Puccetti},
  {Walton}, {Koss}, {Boggs}, \& {Brandt}}]{arevalo14}
{Ar{\'e}valo}, P., {Bauer}, F.~E., {Puccetti}, S., {et~al.} 2014, \apj, 791, 81

\bibitem[{{Assef} {et~al.}(2013){Assef}, {Stern}, {Kochanek}, {Blain},
  {Brodwin}, {Brown}, {Donoso}, {Eisenhardt}, {Jannuzi}, {Jarrett}, {Stanford},
  {Tsai}, {Wu}, \& {Yan}}]{assef13}
{Assef}, R.~J., {Stern}, D., {Kochanek}, C.~S., {et~al.} 2013, \apj, 772, 26

\bibitem[{{Awaki} {et~al.}(2009){Awaki}, {Terashima}, {Higaki}, \&
  {Fukazawa}}]{awaki09}
{Awaki}, H., {Terashima}, Y., {Higaki}, Y., \& {Fukazawa}, Y. 2009, \pasj, 61,
  317

\bibitem[{{Ballantyne} {et~al.}(2014){Ballantyne}, {Bollenbacher}, {Brenneman},
  {Madsen}, {Balokovi{\'c}}, {Boggs}, {Christensen}, {Craig}, {Gandhi},
  {Hailey}, {Harrison}, {Lohfink}, {Marinucci}, {Markwardt}, {Stern}, {Walton},
  \& {Zhang}}]{ballantyne14}
{Ballantyne}, D.~R., {Bollenbacher}, J.~M., {Brenneman}, L.~W., {et~al.} 2014,
  \apj, 794, 62

\bibitem[{{Balokovi{\'c}} {et~al.}(2014){Balokovi{\'c}}, {Comastri},
  {Harrison}, {Alexander}, {Ballantyne}, {Bauer}, {Boggs}, {Brandt},
  {Brightman}, {Christensen}, {Craig}, {Del Moro}, {Gandhi}, {Hailey}, {Koss},
  {Lansbury}, {Luo}, {Madejski}, {Marinucci}, {Matt}, {Markwardt}, {Puccetti},
  {Reynolds}, {Risaliti}, {Rivers}, {Stern}, {Walton}, \&
  {Zhang}}]{balokovic14}
{Balokovi{\'c}}, M., {Comastri}, A., {Harrison}, F.~A., {et~al.} 2014, \apj,
  794, 111

\bibitem[{Bambynek {et~al.}(1972)Bambynek, Crasemann, Fink, Freund, Mark,
  Swift, Price, \& Rao}]{bambynek72}
Bambynek, W., Crasemann, B., Fink, R.~W., {et~al.} 1972, Rev. Mod. Phys., 44,
  716

\bibitem[{{Bauer} {et~al.}(2014){Bauer}, {Arevalo}, {Walton}, {Koss},
  {Puccetti}, {Gandhi}, {Stern}, {Alexander}, {Balokovic}, {Boggs}, {Brandt},
  {Brightman}, {Christensen}, {Comastri}, {Craig}, {Del Moro}, {Hailey},
  {Harrison}, {Hickox}, {Luo}, {Markwardt}, {Marinucci}, {Matt}, {Rigby},
  {Rivers}, {Saez}, {Treister}, {Urry}, \& {Zhang}}]{bauer14}
{Bauer}, F.~E., {Arevalo}, P., {Walton}, D.~J., {et~al.} 2014, ArXiv e-prints,
  arXiv:1411.0670

\bibitem[{{Beckmann} {et~al.}(2009){Beckmann}, {Soldi}, {Ricci},
  {Alfonso-Garz{\'o}n}, {Courvoisier}, {Domingo}, {Gehrels}, {Lubi{\'n}ski},
  {Mas-Hesse}, \& {Zdziarski}}]{beckmann09}
{Beckmann}, V., {Soldi}, S., {Ricci}, C., {et~al.} 2009, \aap, 505, 417

\bibitem[{{Bian} \& {Gu}(2007)}]{bian07}
{Bian}, W., \& {Gu}, Q. 2007, ApJ, 657, 159

\bibitem[{{Brenneman} {et~al.}(2014){Brenneman}, {Madejski}, {Fuerst}, {Matt},
  {Elvis}, {Harrison}, \& {Ballantyne}}]{brenneman14}
{Brenneman}, L.~W., {Madejski}, G., {Fuerst}, F., {et~al.} 2014, \apj, 788, 61

\bibitem[{{Brightman} \& {Nandra}(2011{\natexlab{a}})}]{brightman11}
{Brightman}, M., \& {Nandra}, K. 2011{\natexlab{a}}, \mnras, 413, 1206

\bibitem[{{Brightman} \& {Nandra}(2011{\natexlab{b}})}]{brightman11b}
---. 2011{\natexlab{b}}, \mnras, 414, 3084

\bibitem[{{Brightman} {et~al.}(2014){Brightman}, {Nandra}, {Salvato}, {Hsu},
  {Aird}, \& {Rangel}}]{brightman14}
{Brightman}, M., {Nandra}, K., {Salvato}, M., {et~al.} 2014, \mnras, 443, 1999

\bibitem[{{Brightman} \& {Ueda}(2012)}]{brightman12b}
{Brightman}, M., \& {Ueda}, Y. 2012, \mnras, 423, 702

\bibitem[{{Buchner} {et~al.}(2014){Buchner}, {Georgakakis}, {Nandra}, {Hsu},
  {Rangel}, {Brightman}, {Merloni}, {Salvato}, {Donley}, \&
  {Kocevski}}]{buchner14}
{Buchner}, J., {Georgakakis}, A., {Nandra}, K., {et~al.} 2014, \aap, 564, A125

\bibitem[{{Burlon} {et~al.}(2011){Burlon}, {Ajello}, {Greiner}, {Comastri},
  {Merloni}, \& {Gehrels}}]{burlon11}
{Burlon}, D., {Ajello}, M., {Greiner}, J., {et~al.} 2011, \apj, 728, 58

\bibitem[{{Comastri} {et~al.}(2011){Comastri}, {Ranalli}, {Iwasawa}, {Vignali},
  {Gilli}, {Georgantopoulos}, {Barcons}, {Brandt}, {Brunner}, {Brusa},
  {Cappelluti}, {Carrera}, {Civano}, {Fiore}, {Hasinger}, {Mainieri}, \&
  {Merloni}}]{comastri11}
{Comastri}, A., {Ranalli}, P., {Iwasawa}, K., {et~al.} 2011, \aap, 526, L9+

\bibitem[{{Done} {et~al.}(2003){Done}, {Madejski}, {{\.Z}ycki}, \&
  {Greenhill}}]{done03}
{Done}, C., {Madejski}, G.~M., {{\.Z}ycki}, P.~T., \& {Greenhill}, L.~J. 2003,
  ApJ, 588, 763

\bibitem[{{Draper} \& {Ballantyne}(2010)}]{draper10}
{Draper}, A.~R., \& {Ballantyne}, D.~R. 2010, \apjl, 715, L99

\bibitem[{{Eguchi} {et~al.}(2011){Eguchi}, {Ueda}, {Awaki}, {Aird},
  {Terashima}, \& {Mushotzky}}]{eguchi11}
{Eguchi}, S., {Ueda}, Y., {Awaki}, H., {et~al.} 2011, \apj, 729, 31

\bibitem[{{Elitzur} \& {Shlosman}(2006)}]{elitzur06}
{Elitzur}, M., \& {Shlosman}, I. 2006, \apjl, 648, L101

\bibitem[{{Gandhi} {et~al.}(2009){Gandhi}, {Horst}, {Smette}, {H{\"o}nig},
  {Comastri}, {Gilli}, {Vignali}, \& {Duschl}}]{gandhi09}
{Gandhi}, P., {Horst}, H., {Smette}, A., {et~al.} 2009, \aap, 502, 457

\bibitem[{{Gandhi} {et~al.}(2014){Gandhi}, {Lansbury}, {Alexander}, {Stern},
  {Ar{\'e}valo}, {Ballantyne}, {Balokovi{\'c}}, {Bauer}, {Boggs}, {Brandt},
  {Brightman}, {Christensen}, {Comastri}, {Craig}, {Del Moro}, {Elvis},
  {Fabian}, {Hailey}, {Harrison}, {Hickox}, {Koss}, {LaMassa}, {Luo},
  {Madejski}, {Ptak}, {Puccetti}, {Teng}, {Urry}, {Walton}, \&
  {Zhang}}]{gandhi14}
{Gandhi}, P., {Lansbury}, G.~B., {Alexander}, D.~M., {et~al.} 2014, \apj, 792,
  117

\bibitem[{{Ghisellini} {et~al.}(1994){Ghisellini}, {Haardt}, \&
  {Matt}}]{ghisellini94}
{Ghisellini}, G., {Haardt}, F., \& {Matt}, G. 1994, MNRAS, 267, 743

\bibitem[{{Goulding} {et~al.}(2012){Goulding}, {Alexander}, {Bauer}, {Forman},
  {Hickox}, {Jones}, {Mullaney}, \& {Trichas}}]{goulding12}
{Goulding}, A.~D., {Alexander}, D.~M., {Bauer}, F.~E., {et~al.} 2012, \apj,
  755, 5

\bibitem[{{Greene} {et~al.}(2010){Greene}, {Peng}, {Kim}, {Kuo}, {Braatz},
  {Impellizzeri}, {Condon}, {Lo}, {Henkel}, \& {Reid}}]{greene10}
{Greene}, J.~E., {Peng}, C.~Y., {Kim}, M., {et~al.} 2010, \apj, 721, 26

\bibitem[{{Greenhill} {et~al.}(2003){Greenhill}, {Kondratko}, {Lovell},
  {Kuiper}, {Moran}, {Jauncey}, \& {Baines}}]{greenhill03}
{Greenhill}, L.~J., {Kondratko}, P.~T., {Lovell}, J.~E.~J., {et~al.} 2003,
  \apjl, 582, L11

\bibitem[{{Greenhill} {et~al.}(1997){Greenhill}, {Moran}, \&
  {Herrnstein}}]{greenhill97}
{Greenhill}, L.~J., {Moran}, J.~M., \& {Herrnstein}, J.~R. 1997, \apjl, 481,
  L23

\bibitem[{{Gu}(2013)}]{gu13}
{Gu}, M. 2013, \apj, 773, 176

\bibitem[{{Guainazzi} \& {Bianchi}(2007)}]{guainazzi07}
{Guainazzi}, M., \& {Bianchi}, S. 2007, \mnras, 374, 1290

\bibitem[{{Harrison} {et~al.}(2013){Harrison}, {Craig}, {Christensen},
  {Hailey}, \& {Zhang}}]{harrison13}
{Harrison}, F.~A., {Craig}, W.~W., {Christensen}, F.~E., {Hailey}, C.~J., \&
  {Zhang}, W.~W. 2013, \apj, 770, 103

\bibitem[{{Hasinger}(2008)}]{hasinger08}
{Hasinger}, G. 2008, \aap, 490, 905

\bibitem[{{Ikeda} {et~al.}(2009){Ikeda}, {Awaki}, \& {Terashima}}]{ikeda09}
{Ikeda}, S., {Awaki}, H., \& {Terashima}, Y. 2009, \apj, 692, 608

\bibitem[{{Ishihara} {et~al.}(2001){Ishihara}, {Nakai}, {Iyomoto}, {Makishima},
  {Diamond}, \& {Hall}}]{ishihara01}
{Ishihara}, Y., {Nakai}, N., {Iyomoto}, N., {et~al.} 2001, \pasj, 53, 215

\bibitem[{{Kawamuro} {et~al.}(2013){Kawamuro}, {Ueda}, {Tazaki}, \&
  {Terashima}}]{kawamuro13}
{Kawamuro}, T., {Ueda}, Y., {Tazaki}, F., \& {Terashima}, Y. 2013, \apj, 770,
  157

\bibitem[{{Kondratko} {et~al.}(2005){Kondratko}, {Greenhill}, \&
  {Moran}}]{kondratko05}
{Kondratko}, P.~T., {Greenhill}, L.~J., \& {Moran}, J.~M. 2005, \apj, 618, 618

\bibitem[{{La Franca} {et~al.}(2005){La Franca}, {Fiore}, {Comastri}, {Perola},
  {Sacchi}, {Brusa}, {Cocchia}, {Feruglio}, {Matt}, {Vignali}, {Carangelo},
  {Ciliegi}, {Lamastra}, {Maiolino}, {Mignoli}, {Molendi}, \&
  {Puccetti}}]{lafranca05}
{La Franca}, F., {Fiore}, F., {Comastri}, A., {et~al.} 2005, \apj, 635, 864

\bibitem[{{Lawrence}(1991)}]{lawrence91}
{Lawrence}, A. 1991, \mnras, 252, 586

\bibitem[{{Lawrence} \& {Elvis}(2010)}]{lawrence10}
{Lawrence}, A., \& {Elvis}, M. 2010, \apj, 714, 561

\bibitem[{{Leahy} \& {Creighton}(1993)}]{leahy93}
{Leahy}, D.~A., \& {Creighton}, J. 1993, MNRAS, 263, 314

\bibitem[{{Liu} \& {Li}(2014)}]{liu14}
{Liu}, Y., \& {Li}, X. 2014, \apj, 787, 52

\bibitem[{{Liu} \& {Li}(2015)}]{liu15}
---. 2015, \mnras, 448, L53

\bibitem[{{Lusso} {et~al.}(2013){Lusso}, {Hennawi}, {Comastri}, {Zamorani},
  {Richards}, {Vignali}, {Treister}, {Schawinski}, {Salvato}, \&
  {Gilli}}]{lusso13}
{Lusso}, E., {Hennawi}, J.~F., {Comastri}, A., {et~al.} 2013, \apj, 777, 86

\bibitem[{{Madejski} {et~al.}(2000){Madejski}, {{\.Z}ycki}, {Done}, {Valinia},
  {Blanco}, {Rothschild}, \& {Turek}}]{madejski00}
{Madejski}, G., {{\.Z}ycki}, P., {Done}, C., {et~al.} 2000, \apjl, 535, L87

\bibitem[{{Magdziarz} \& {Zdziarski}(1995)}]{magdziarz95}
{Magdziarz}, P., \& {Zdziarski}, A.~A. 1995, \mnras, 273, 837

\bibitem[{{Marinucci} {et~al.}(2012){Marinucci}, {Bianchi}, {Nicastro}, {Matt},
  \& {Goulding}}]{marinucci12}
{Marinucci}, A., {Bianchi}, S., {Nicastro}, F., {Matt}, G., \& {Goulding},
  A.~D. 2012, \apj, 748, 130

\bibitem[{{Marinucci} {et~al.}(2014){Marinucci}, {Matt}, {Kara}, {Miniutti},
  {Elvis}, {Arevalo}, {Ballantyne}, {Balokovi{\'c}}, {Bauer}, {Brenneman},
  {Boggs}, {Cappi}, {Christensen}, {Craig}, {Fabian}, {Fuerst}, {Hailey},
  {Harrison}, {Risaliti}, {Reynolds}, {Stern}, {Walton}, \&
  {Zhang}}]{marinucci14}
{Marinucci}, A., {Matt}, G., {Kara}, E., {et~al.} 2014, \mnras, 440, 2347

\bibitem[{{Markowitz} {et~al.}(2014){Markowitz}, {Krumpe}, \&
  {Nikutta}}]{markowitz14}
{Markowitz}, A.~G., {Krumpe}, M., \& {Nikutta}, R. 2014, \mnras, 439, 1403

\bibitem[{{Matt} {et~al.}(1991){Matt}, {Perola}, \& {Piro}}]{matt91}
{Matt}, G., {Perola}, G.~C., \& {Piro}, L. 1991, \aap, 247, 25

\bibitem[{{Mayo} \& {Lawrence}(2013)}]{mayo13}
{Mayo}, J.~H., \& {Lawrence}, A. 2013, \mnras, 434, 1593

\bibitem[{{Molendi} {et~al.}(2003){Molendi}, {Bianchi}, \& {Matt}}]{molendi03}
{Molendi}, S., {Bianchi}, S., \& {Matt}, G. 2003, \mnras, 343, L1

\bibitem[{{Murphy} \& {Yaqoob}(2009)}]{murphy09}
{Murphy}, K.~D., \& {Yaqoob}, T. 2009, \mnras, 397, 1549

\bibitem[{{Nandra} \& {George}(1994)}]{nandra94monte}
{Nandra}, K., \& {George}, I.~M. 1994, MNRAS, 267, 974

\bibitem[{{Puccetti} {et~al.}(2014){Puccetti}, {Comastri}, {Fiore},
  {Ar{\'e}valo}, {Risaliti}, {Bauer}, {Brandt}, {Stern}, {Harrison},
  {Alexander}, {Boggs}, {Christensen}, {Craig}, {Gandhi}, {Hailey}, {Koss},
  {Lansbury}, {Luo}, {Madejski}, {Matt}, {Walton}, \& {Zhang}}]{puccetti14}
{Puccetti}, S., {Comastri}, A., {Fiore}, F., {et~al.} 2014, \apj, 793, 26

\bibitem[{{Risaliti} {et~al.}(1999){Risaliti}, {Maiolino}, \&
  {Salvati}}]{risaliti99}
{Risaliti}, G., {Maiolino}, R., \& {Salvati}, M. 1999, ApJ, 522, 157

\bibitem[{{Sambruna} {et~al.}(2001){Sambruna}, {Netzer}, {Kaspi}, {Brandt},
  {Chartas}, {Garmire}, {Nousek}, \& {Weaver}}]{sambruna11}
{Sambruna}, R.~M., {Netzer}, H., {Kaspi}, S., {et~al.} 2001, \apjl, 546, L13

\bibitem[{{Simpson}(2005)}]{simpson05}
{Simpson}, C. 2005, \mnras, 360, 565

\bibitem[{{Su} {et~al.}(2008){Su}, {Zhang}, \& {Fan}}]{su08}
{Su}, J.-B., {Zhang}, J.-S., \& {Fan}, J.-H. 2008, \cjaa, 8, 547

\bibitem[{{Tazaki} {et~al.}(2011){Tazaki}, {Ueda}, {Terashima}, \&
  {Mushotzky}}]{tazaki11}
{Tazaki}, F., {Ueda}, Y., {Terashima}, Y., \& {Mushotzky}, R.~F. 2011, \apj,
  738, 70

\bibitem[{{Tilak} {et~al.}(2008){Tilak}, {Greenhill}, {Done}, \&
  {Madejski}}]{tilak08}
{Tilak}, A., {Greenhill}, L.~J., {Done}, C., \& {Madejski}, G. 2008, \apj, 678,
  701

\bibitem[{{Treister} {et~al.}(2008){Treister}, {Krolik}, \&
  {Dullemond}}]{treister08}
{Treister}, E., {Krolik}, J.~H., \& {Dullemond}, C. 2008, \apj, 679, 140

\bibitem[{{Tueller} {et~al.}(2008){Tueller}, {Mushotzky}, {Barthelmy},
  {Cannizzo}, {Gehrels}, {Markwardt}, {Skinner}, \& {Winter}}]{tueller08}
{Tueller}, J., {Mushotzky}, R.~F., {Barthelmy}, S., {et~al.} 2008, \apj, 681,
  113

\bibitem[{{Ueda} {et~al.}(2003){Ueda}, {Akiyama}, {Ohta}, \& {Miyaji}}]{ueda03}
{Ueda}, Y., {Akiyama}, M., {Ohta}, K., \& {Miyaji}, T. 2003, ApJ, 598, 886

\bibitem[{{Vasudevan} {et~al.}(2013){Vasudevan}, {Brandt}, {Mushotzky},
  {Winter}, {Baumgartner}, {Shimizu}, {Schneider}, \& {Nousek}}]{vasudevan13}
{Vasudevan}, R.~V., {Brandt}, W.~N., {Mushotzky}, R.~F., {et~al.} 2013, \apj,
  763, 111

\bibitem[{{Verner} {et~al.}(1996){Verner}, {Ferland}, {Korista}, \&
  {Yakovlev}}]{verner96}
{Verner}, D.~A., {Ferland}, G.~J., {Korista}, K.~T., \& {Yakovlev}, D.~G. 1996,
  ApJ, 465, 487

\bibitem[{{Walton} {et~al.}(2013){Walton}, {Fuerst}, {Harrison}, {Stern},
  {Bachetti}, {Barret}, {Bauer}, {Boggs}, {Christensen}, {Craig}, {Fabian},
  {Grefenstette}, {Hailey}, {Madsen}, {Miller}, {Ptak}, {Rana}, {Webb}, \&
  {Zhang}}]{walton13}
{Walton}, D.~J., {Fuerst}, F., {Harrison}, F., {et~al.} 2013, \apj, 779, 148

\bibitem[{{Walton} {et~al.}(2014){Walton}, {Harrison}, {Grefenstette},
  {Miller}, {Bachetti}, {Barret}, {Boggs}, {Christensen}, {Craig}, {Fabian},
  {Fuerst}, {Hailey}, {Madsen}, {Parker}, {Ptak}, {Rana}, {Stern}, {Webb}, \&
  {Zhang}}]{walton14}
{Walton}, D.~J., {Harrison}, F.~A., {Grefenstette}, B.~W., {et~al.} 2014, \apj,
  793, 21

\bibitem[{{Yaqoob}(1997)}]{yaqoob97}
{Yaqoob}, T. 1997, ApJ, 479, 184

\bibitem[{{Yaqoob}(2012)}]{yaqoob12}
---. 2012, \mnras, 423, 3360

\end{thebibliography}

\end{document}